\documentclass[preprint, amsmath,amssymb, aps, physrev]{revtex4-2}
\usepackage{subfigure,dcolumn}
\usepackage[T2A,T1]{fontenc}
\usepackage[russian,english]{babel}

\usepackage{listings}
\usepackage[utf8]{inputenc}
\usepackage{CJK}
\usepackage{kotex}
\usepackage{graphicx}
\usepackage{amssymb}
\usepackage{amsmath,amsthm,latexsym,amsfonts}
\usepackage{booktabs}
\usepackage{rotating}
\usepackage{multirow}
\usepackage{tablefootnote}
\usepackage{array}
\usepackage{siunitx}
\usepackage{hhline}
\usepackage{tabularx}
\usepackage{romannum}
\usepackage{mathrsfs}
\usepackage{longtable}
\usepackage{dcolumn}
\usepackage{bm}
\usepackage{setspace}
\usepackage{xcolor}

\AtBeginDocument{\pagenumbering{arabic}}
\begin{document}
\preprint{APS/123-QED}
\setlength{\LTcapwidth}{\textwidth}

\title{\textbf{Systematic Bayesian Evaluation of Resonance Parameters in $^{19}$Ne for the $^{15}$O($\alpha, \gamma$)$^{19}$Ne and $^{18}$F($p, \alpha$)$^{15}$O Reactions}}

\author{S.~H.~\surname{Kim} (김소현)}
\affiliation{Department of Physics, Sungkyunkwan University, Suwon 16419, Republic of Korea}
\email{sohyun114@skku.edu}

\author{K.~Y.~\surname{Chae} (채경육)}
\email{kchae@skku.edu}
\thanks{Fax: +82-31-290-7055}
\affiliation{Department of Physics, Sungkyunkwan University, Suwon 16419, Republic of Korea}

\author{C.~H.~\surname{Kim} (김찬희)}
\affiliation{Department of Physics, Sungkyunkwan University, Suwon 16419, Republic of Korea}
\email{chkim@phys.kim}

\author{{C. D. \surname{Nesaraja}}}
\affiliation{Physics Division, Oak Ridge National Laboratory, Oak Ridge, TN 37831, USA}

\author{{M. S. \surname{Smith}}}
\affiliation{Physics Division, Oak Ridge National Laboratory, Oak Ridge, TN 37831, USA}
\affiliation{Stellar Science Solutions, Lenoir City, TN 37772, USA}

\date{\today}
\begin{abstract}


We present a comprehensive evaluation of the nuclear structure properties of $^{19}$Ne using a novel and rigorous Bayesian statistical framework. Precise characterization of $^{19}$Ne resonance parameters is critical for accurately determining reaction rates of the astrophysically significant $^{15}$O($\alpha, \gamma$)$^{19}$Ne and $^{18}$F($p, \alpha$)$^{15}$O reactions, which govern breakout from the hot CNO cycle in X-ray bursts and influence gamma-ray emission in novae, respectively. By reconstructing likelihood functions from published experimental data -- including asymmetric uncertainties and upper or lower limits -- we derive posterior distributions for resonance energies, decay widths, and branching ratios. Our Bayesian approach systematically incorporates previously reported discrepancies among measurements, providing a statistically robust and consistent treatment of these uncertainties. The evaluated resonance parameters and associated uncertainties provide crucial input for stellar nucleosynthesis modeling, contributing to a refined understanding of explosive astrophysical phenomena.

\end{abstract}

\maketitle

\section{\label{sec:level1}Introduction}

$^{19}$Ne is a critical nuclide in astrophysics since it is the compound nucleus of the $^{15}$O($\alpha, \gamma$)$^{19}$Ne and the $^{18}$F($p, \alpha$)$^{15}$O reactions. The $^{15}$O($\alpha$, $\gamma$)$^{19}$Ne reaction enables the breakout from the hot CNO (HCNO) cycle at temperatures above 0.4–0.5 GK in X-ray bursts (XRBs)\cite{1981Wa}. When its rate exceeds the $\beta$-decay rate of $^{15}$O, $^{19}$Ne is produced and rapidly converted to $^{20}$Na via proton capture, triggering greatly enhanced nuclear energy production via the rapid proton-capture ($rp$) process. As such, the $^{15}$O($\alpha$, $\gamma$)$^{19}$Ne rate critically affects XRB dynamics, influencing light curves and nucleosynthesis~\cite{2006Fi,2016Cy}. The $^{18}$F($p, \alpha$)$^{15}$O reaction destroys $^{18}$F, which is a key source of the 511~keV $\gamma$-rays from positron annihilation in novae~\cite{1974Cl,1987Le,1999He}. The $^{18}$F abundance, which determines the strength of the emission, is most strongly affected by the $^{18}$F($p, \alpha$)$^{15}$O reaction among all production and destruction channels~\cite{2002Ii,2000Co,1999He}.

The level structure of $^{19}$Ne has been extensively investigated to reduce uncertainties in the $^{15}$O($\alpha$, $\gamma$)$^{19}$Ne and $^{18}$F($p, \alpha$)$^{15}$O reaction rates at stellar temperatures. Various experimental approaches have been employed to measure properties of the relevant resonance levels, providing critical inputs for accurate thermonuclear reaction rate calculations. Based on these experimental results, numerous studies have been conducted to quantify these rates. For example, Illiadis~\textit{et al.} \cite{2010Ii} calculated the $^{15}$O($\alpha$, $\gamma$)$^{19}$Ne rate using a Monte Carlo method combined with the TALYS code~\cite{2010Ii}, and Davids~\textit{et al.} \cite{2011Da} investigated the impact of rate uncertainties on XRB luminosity, isotopic abundances, and peak temperatures. Nesaraja~\textit{et al.} \cite{2007Ne} compiled nuclear data for $^{19}$Ne levels above the proton threshold and predicted missing states by comparison with its mirror nucleus $^{19}$F. Kahl~\textit{et al.} \cite{2021Ka} reviewed recent measurements and examined the influence of the $^{18}$F($p, \alpha$)$^{15}$O rate on nucleosynthesis in novae.

The aforementioned studies provide valuable guidance in identifying resonance levels that require further investigation to reduce reaction rate uncertainties. However, they typically rely on a limited subset of experimental data -- primarily the most recent and reliable measurements -- as input for rate calculations. This selective approach may lead to underestimation or overestimation of uncertainties, especially when discrepancies exist among the experimental results. While weighted averaging is often used to obtain a recommended value for measurements of a physical quantity, it may become invalid when the likelihoods deviate significantly from a Gaussian form. Therefore, it is important to consider statistical methods that can be broadly applicable to nuclear data evaluations.

In this work, we use a novel approach wherein nuclear structure properties of $^{19}$Ne levels are systematically combined from existing experimental measurements using a Bayesian approach, and likelihoods are reconstructed directly from the reported values. The basic concept of Bayes' theorem is introduced in Sec.~\ref{sec:Bayes Theorem}, followed by a discussion of the prior distributions for each resonance parameter in Sec.~\ref{sec:prior}. The likelihood function is reconstructed using statistical approximations and theoretical considerations as described in Sec.~\ref{sec:likelihood}. For the cases where the Bayesian approach is challenging, additional approaches -- such as spin-parity ($J^{\pi}$) assignments and mirror-state analyses -- are applied and presented in Sec.~\ref{sec:spin} and Sec.~\ref{sec: mirror}. In addition, the statistical conversion from mean lifetime $\tau_m$ to decay width $\Gamma$ is explained in Sec.~\ref{sec: tau_gamma}. The level structure of $^{19}$Ne is evaluated using the proposed methodology and available experimental data in Sec.~\ref{sec: level information}.

\section{\label{sec:Method}Method}
\subsection{\label{sec:Bayes Theorem}Bayes' Theorem}
Resonance parameters, such as excitation energy E$_x$, decay width $\Gamma$, charged-particle decay width $\Gamma_\lambda$, and branching ratio $B_\alpha$, of $^{19}$Ne levels obtained from independent measurements are evaluated in this work using Bayes' theorem. When a data set $\textit{X}$ is measured for a model parameter $\theta$, Bayes’ theorem can be expressed as  

\begin{align}
\textit{p}(\theta\mid{X})=\frac{\textit{P}(X\mid\theta)\pi(\theta)}{\int\textit{P}(X\mid\theta)\pi(\theta)d\theta}=\frac{\textit{L}(\theta\mid{X})\pi(\theta)}{\int\textit{L}(\theta\mid{X})\pi(\theta)d\theta},
\label{eqn: Bayes}
\end{align}

\noindent where $\pi(\theta)$ is the prior, $P(X\mid\theta)=L(\theta \mid X)$ is the likelihood, and $p(\theta \mid X)$ is the posterior. The prior represents existing knowledge or assumptions about the parameter $\theta$ before new data are taken into account. The likelihood represents the probability of obtaining the observed dataset $X$ given a specific parameter value $\theta$. The standard notation of likelihood is $P(X\mid\theta)$, but $L(\theta \mid{X})$ is used in the present work to emphasize that it is a function of the parameter $\theta$ as previously done in Refs.~\cite{Bayes_1,Bayes_2,Bayes_3}. The posterior reflects the updated knowledge about $\theta$ after incorporating the data. The denominator in Eq.~(\ref{eqn: Bayes}), referred to as the evidence, serves as a normalization factor.

Bayes' theorem provides a formal framework for updating prior knowledge about model parameters by incorporating new observational data. The resulting posterior can subsequently serve as a new prior when additional data becomes available. By repeating this process, if $N$ independent measurements are performed for a parameter $\theta$, with $L_i(\theta \mid X_i)$ denoting the likelihood from the $i$th measurement and $\pi(\theta)$ the prior, the final posterior is given by:

\begin{align}
&p_{final}(\theta\mid{X_1, X_2,\cdots,X_N})\notag \\
&=\dfrac{L_1(\theta \mid{X_1})L_2(\theta\mid{X_2})\cdots L_N(\theta\mid{X_N})\pi(\theta)}{\int L_1(\theta\mid{X_1})\int L_2(\theta\mid{X_2})\cdots \int L_N(\theta\mid{X_N})\pi(\theta)d^N\theta} \notag \\
&=\dfrac{\prod_{i=1}^{N}L_i(\theta\mid{X_i})\pi(\theta)}{ \prod_{i=1}^{N}\int_i L_i(\theta\mid{X_i})\pi(\theta)d_i\theta },
\label{eqn:evaluate_parameters}
\end{align}

\noindent where $\textit{X}$ represents data obtained from independent experiments and $\theta$ denotes a resonance parameter such as the excitation energy E$_x$, alpha branching ratio B$_\alpha$, or total decay width $\Gamma$.

In principle, the likelihood is obtained from the raw experimental data $\textit{X}$ and requires complex analysis. However, if the likelihood can be reconstructed from reported results for the parameter $\theta$, the evaluation of parameters becomes simple based on Eq.~(\ref{eqn:evaluate_parameters}). Incorporating new measurement results is also straightforward, requiring only the multiplication of the likelihood of new data with the previous posterior.


\subsection{\label{sec:prior}Choice of prior}
Given the importance of prior selection, several prior distributions are considered. Each prior is constructed based on established nuclear physics principles relevant to the corresponding resonance parameter. The prior selection for each resonance parameter is summarized in Table~\ref{tab:priors}.

\begin{figure*}
\centering

\includegraphics[width = 0.8\textwidth]{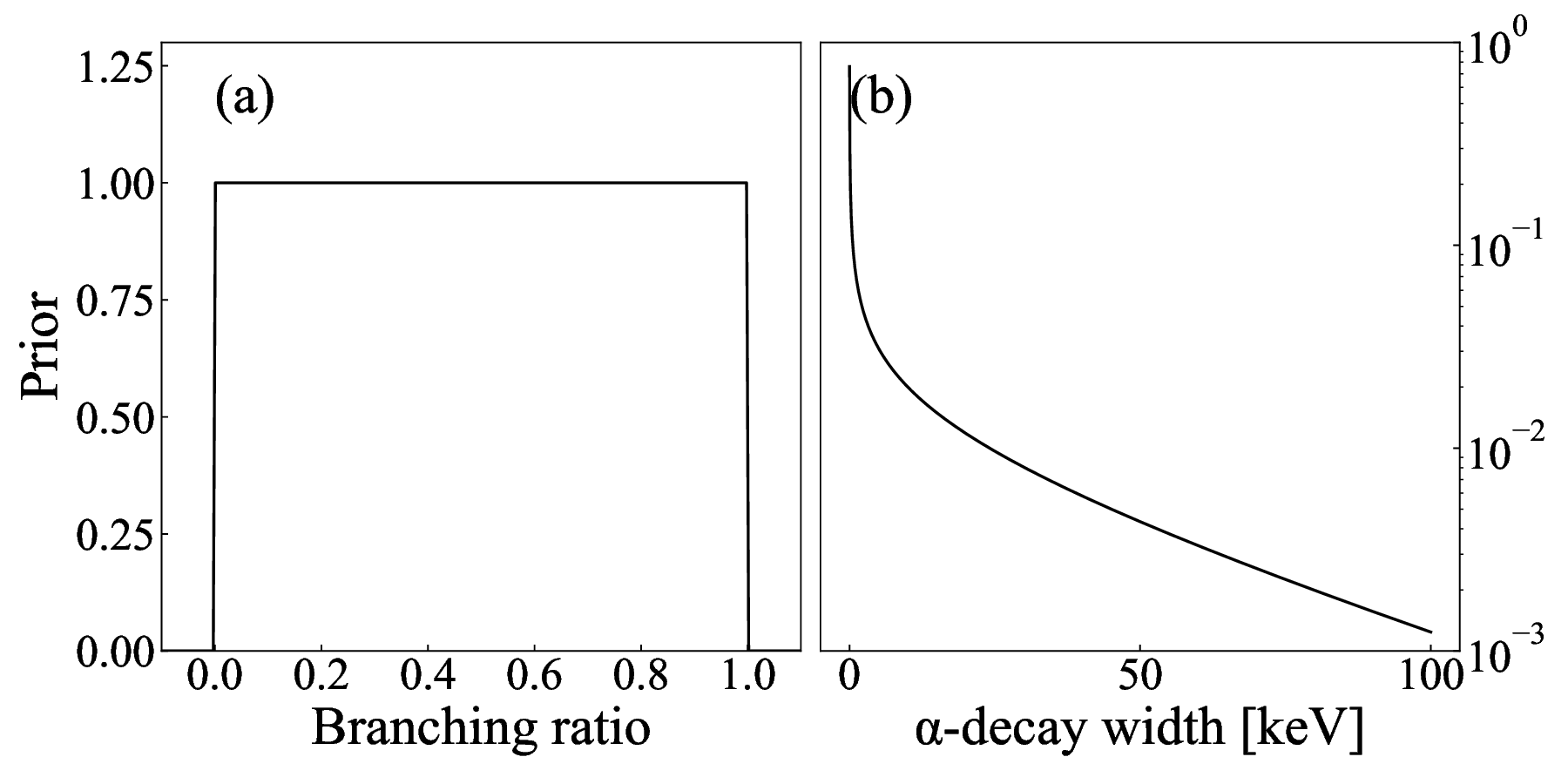}
\caption{Examples of priors for (a) branching ratio and (b) $\alpha$-decay width.}
\label{fig: example_prior}

\end{figure*}

\begin{table}[htp!]
    \centering
    \caption{ \label{tab:priors}Summary of prior selections for each resonance parameter.}
    \begin{ruledtabular}

    \begin{tabular}{cc}
    
         Parameter $\theta$&  Priors $\pi(\theta)$\\
         \hline
        $B_\alpha$ &  \parbox{4cm}{
\begin{align*}
\pi(B_\lambda) = 
\left\{
\begin{array}{rcl}
1 & \text{for} & 0 \leq B_\lambda \leq 1 \\
0 & \text{for} & B_\lambda < 0 \text{ or } B_\lambda > 1
\end{array}
\right.
\end{align*}} \\
        $\Gamma_\alpha$ and $\Gamma_p$ & \parbox{4cm}{\begin{align*}
\pi(\Gamma_\lambda) = \dfrac{1}{\sqrt{2\pi\langle\theta^2\rangle A_\lambda \Gamma_\lambda}}e^{-\frac{\Gamma_\lambda}{2\langle\theta^2\rangle A_\lambda}}
\end{align*}} \\
        $\Gamma$ and E$_x$ & \parbox{4cm}{\begin{align*}
\pi(\theta) = \Theta(\theta) =
\left\{
\begin{array}{rcl}
1 & \mbox{for} & 0 \leq \theta \\
0 & \mbox{for} & 0 > \theta
\end{array}
\right.
\end{align*}}\\
    \end{tabular}
    \end{ruledtabular}
\end{table}

The branching ratio for a specific decay channel $\lambda$ must satisfy $0 \leq B_\lambda \leq 1$ by definition. Therefore, a step function is a natural choice for its prior distribution:

\begin{align}
\pi(B_\lambda) = 
\left\{
\begin{array}{rcl}
1 & \mbox{for} & 0 \leq B_\lambda \leq 1 \\
0 & \mbox{for} & B_\lambda < 0 \ \mbox{or} \ B_\lambda > 1.
\end{array}
\right.
\label{eqn: branching ratio prior}
\end{align}

\noindent This prior assigns equal probability to all physically allowed values and excludes regions not allowed.

For a given charged-particle decay channel $\lambda$, the prior for partial decay width $\Gamma_\lambda$ is derived from the Porter-Thomas distribution~\cite{PT_dist,2010Lo,NiCu_eval}. The distribution of the dimensionless variable $\theta^2_\lambda / \langle \theta^2_\lambda \rangle$, where $\theta^2_\lambda$ is the reduced partial width, follows a first-order chi-square distribution~\cite{PT_dist}:

\begin{align}
\pi(\theta^2_\lambda) = \dfrac{1}{\sqrt{2\pi\langle\theta^2_\lambda\rangle\theta^2_\lambda}}e^{-\frac{\theta^2_\lambda}{2\langle\theta^2_\lambda\rangle}}.
\label{eqn: reduced_PT}
\end{align}

\noindent This expression can be transformed to the distribution of $\Gamma_\lambda$ using the relation
$\Gamma_\lambda = \dfrac{2\hbar P_{l\lambda}}{\mu R^2}\theta^2_\lambda = A_\lambda \theta^2_\lambda$, where $\mu$ is the reduced mass, $P_{l\lambda}$ is the penetration factor for decay channel $\lambda$ with orbital angular momentum $l$, and $R$ is the interaction radius. The resulting prior becomes:


\begin{align}
\pi(\Gamma_\lambda) = \dfrac{1}{\sqrt{2\pi\langle\theta_\lambda^2\rangle A_\lambda \Gamma_\lambda}}e^{-\frac{\Gamma_\lambda}{2\langle\theta_\lambda^2\rangle A_\lambda}}.
\label{eqn: Gamma_PT}
\end{align}


\noindent In this work, the mean reduced partial width $\langle \theta^2_\lambda \rangle$ are adopted from Ref.~\cite{2007Ne}: $\langle \theta^2_\alpha\rangle=0.05\pm$0.04, and $\langle \theta^2_p \rangle = 0.1 \pm 0.1$ for positive parity and $0.01 \pm 0.01$ for negative parity~\cite{p_reduced_width,2000Co,1998Uk}. Figure~\ref{fig: example_prior} shows the prior distributions for the branching ratio and the $\alpha$-decay width.

For other positive parameters, such as excitation energy E$_x$ and total decay width $\Gamma$, that lack suitable theoretical or phenomenological constraints, the Heaviside function,

\begin{align}
\pi(\theta) = \Theta(\theta) =
\left\{
\begin{array}{rcl}
1 & \mbox{for} & 0 \leq \theta \\
0 & \mbox{for} & 0 > \theta
\end{array}
\right.
\label{eqn: heaviside}
\end{align}

\noindent is adopted. This reflects the minimal prior knowledge that the parameter must be positive. While the uniform priors in Eqs.~(\ref{eqn: branching ratio prior}) and (\ref{eqn: heaviside}) may appear to have no particular effects on the likelihood, the truncations at the boundaries ensure physically consistent parameter estimation, which will also help Monte Carlo reaction rate calculations. 

\subsection{\label{sec:likelihood}Reconstructing likelihood}

The likelihood is typically inferred from raw experimental data as mentioned in Sec.\ref{sec:Bayes Theorem}. However, by applying the quadratic approximation in conjunction with Wilks' theorem~\cite{Wilks}, it becomes possible to approximate the likelihood function using published values and associated uncertainties. Table~\ref{tab:Likelihood} shows a summary of likelihood distributions reconstructed from various cases of reported parameters. This section describes the definition of the confidence interval derived from Wilks' theorem, and outlines how to reconstruct the likelihood for each case accordingly.

\begin{table*}
\caption{\label{tab:Likelihood}Summary of reconstructed likelihood distributions $L(\theta).$}
\begin{ruledtabular}
\begin{tabular}{cc}
reported $\theta$& Reconstructed likelihood $L(\theta)\propto$ \\
\hline
\vspace{0.5cm}
$\theta_0\pm\sigma$& $ \exp\left[-\dfrac{(\theta-\theta_0)^2}{2\sigma^2}\right]$ \\
\vspace{0.5cm}
${\theta_0}^{+\sigma_1}_{-\sigma_2}$& $ \exp\left[-\dfrac{(\theta-\theta_0)^2}{4}\left(\dfrac{1}{\{\sigma+\sigma'(\theta-\theta_0)\}^2}+\dfrac{1}{V+V'(\theta-\theta_0)}\right)\right]$\footnotemark[1]\\
$\theta<\theta_{up}$ (or $\theta>\theta_{lo})$& $\left\{
\begin{array}{ll}
\text{const.} & \theta \leq \theta_0~(\theta>\theta_0) \\
\exp\left[-\dfrac{(\theta - \theta_0)^2}{2\sigma^2}\right] & \theta > \theta_0~(\theta\leq\theta_0)
\end{array}
\right.$\footnotemark[2] \\
\end{tabular}
\end{ruledtabular}
\footnotemark[1]{$\sigma=\frac{2\sigma_1\sigma_2}{\sigma_1+\sigma_2}, \sigma'=\frac{\sigma_1-\sigma_2}{\sigma_1+\sigma_2}, V = \sigma_1\cdot\sigma_2,$ and $V'=\sigma_1-\sigma_2$.}\\
\footnotemark[2]{$\theta_{up}=\theta_0+1.282\sigma=1.128\theta_0~(\theta_{lo}=\theta_0-1.282\sigma=0.872\theta_0$), assuming $\sigma=0.1\times\theta_0$.} 
\end{table*}

\subsubsection{\label{sec:Wilks theorem}Definition of Confidence Intervals}

A standard method for defining the confidence interval boundaries of a parameter $\theta$ is the likelihood ratio test. It requires a test statistic $t_\theta$ defined as follows to quantify the discrepancy between the experimental data $X$ and the parameter $\theta$:

\begin{align}
t_\theta = -2\Delta\mathscr{L}(\theta)&= -2 \left( \mathscr{L}(\theta \mid X) - \mathscr{L}(\theta_{\text{MLE}} \mid X) \right)\nonumber \\
&= -2 \ln \left( \dfrac{L(\theta \mid X)}{L(\theta_{\text{MLE}} \mid X)} \right).
\label{eqn: t_th}
\end{align}
\noindent Here, $\mathscr{L}(\theta \mid X)$ is the logarithm of the likelihood function and $\theta_{\text{MLE}}$ is the Maximum Likelihood Estimator (MLE). The minimum value of zero occurs at $\theta=\theta_{\text{MLE}}$. The larger value of $t_\theta$ indicates greater incompatibility between the parameter $\theta$ and the observed data. Accordingly, a parameter value $\theta=\theta_{\text{MLE}}+\Delta$ defines the boundary of the confidence interval when the corresponding test statistic, $t_{\theta_{\text{MLE}}+\Delta}$, reaches a specified threshold. 

\begin{figure}
\includegraphics[width = 0.7\textwidth]{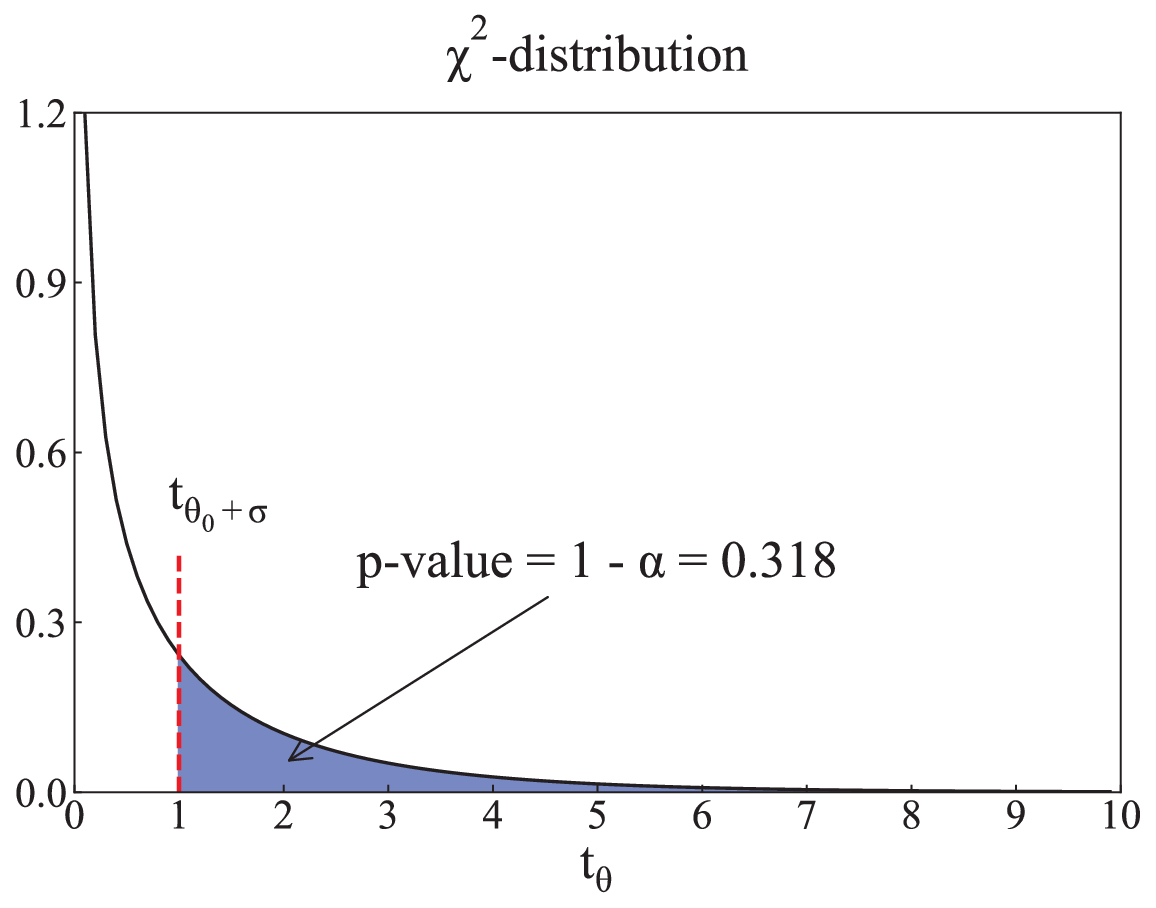}
\caption{The relationship between probability density function of test statistics t$_\theta$, $p$-value, and confidence level $\alpha$ is shown in the figure.}
\label{fig: test_statistics}
\end{figure}

The threshold is determined according to Wilks' theorem~\cite{Wilks}: when the sample size is large, the distribution of $t_\theta$ approaches a chi-square distribution with $k$ degrees of freedom $\chi^2_k(t_\theta)$. The degree of freedom is $k = 1$, since only a single parameter is considered in this work. The level of incompatibility is quantified by the $p$-value, as shown in Fig.~\ref{fig: test_statistics}:

\begin{align}
p_\theta = \int_{t_{\theta_{\text{MLE}}+\Delta}}^{\infty} \chi^2(t_\theta) ~dt_\theta.
\label{eqn: pvalue}
\end{align}

\noindent It represents the probability of obtaining a test statistic equal to or greater than the $t_{\theta_\text{MLE}+\Delta}$. A smaller $p$-value implies less compatibility between the observed data $X$ and the parameter $\theta$. If the $p$-value for $\theta_{\text{MLE}} + \Delta$ is equal to a significance level $\alpha$, the corresponding confidence level will be $1 - \alpha$. Further discussion can be found in Ref.~\cite{2011Co}. At a confidence level of $68.2\%$ (i.e. $\alpha=0.318$), the corresponding threshold for $t_\theta$ is 1 as illustrated in Fig.~\ref{fig: test_statistics}. Accordingly, the boundary of the 68.2\% confidence interval, $\theta_{\text{MLE}}+\sigma$, is defined by the following condition:

\begin{align}
t_{\theta_{\text{MLE}} + \sigma}=1=-2(\mathscr{L}(\theta_{\text{MLE}}+\sigma\mid{X})-\mathscr{L}(\theta_{\text{MLE}}\mid{X})).
\label{eqn: t_1}
\end{align}

\subsubsection{\label{sec: symmetric}Symmetric uncertainty}

When a parameter $\theta$ is reported as $\theta_0\pm\sigma$, it is conventionally understood to indicate that $\theta_0$ represents $\theta_{\text{MLE}}$, and $\sigma$ corresponds to a 68.2\% confidence interval. By the definition of confidence interval in Eq.~(\ref{eqn: t_1}), the likelihood function should satisfy the following condition:

\begin{align}
t_{\theta\pm\sigma}&=-2 \left( \mathscr{L}(\theta_0\pm\sigma \mid X) - \mathscr{L}(\theta_0\mid X) \right) \nonumber \\
&= -2\Delta\mathscr{L}(\theta_0\pm\sigma) = 1.
\label{eqn: symmetric}
\end{align}

\noindent Given that $-2\Delta\mathscr{L}(\theta_0)=0$, the log-likelihood ratio curve $-2\Delta\mathscr{L}(\theta)$ can be approximated by a second-order Taylor expansion around $\theta=\theta_0$:

\begin{align}
\mathscr{L}(\theta) \approx \mathscr{L}(\theta_0) + \frac{1}{2}\ddot{\mathscr{L}}(\theta_0)(\theta-\theta_0)^2.
\label{eqn: Taylor}
\end{align}

\noindent Consequently, using Eqs.~(\ref{eqn: symmetric}) and (\ref{eqn: Taylor}), the likelihood function $L(\theta \mid X)$ can be approximated by a Gaussian distribution centered at $\theta_0$ with standard deviation $\sigma$. Figure~\ref{fig: log-likelihood_example_0} (a) shows the reconstructed $-2\Delta \mathscr{L}$ curves and $L(B_\alpha)$ based on the reported $\alpha-$branching ratios $B_\alpha$ from Ref.~\cite{2009Ta}, as an example.

\subsubsection{\label{sec: asymmetric} Asymmetric uncertainty}

\begin{figure}

\begin{subfigure}
\centering
\includegraphics[width = 0.7\textwidth]{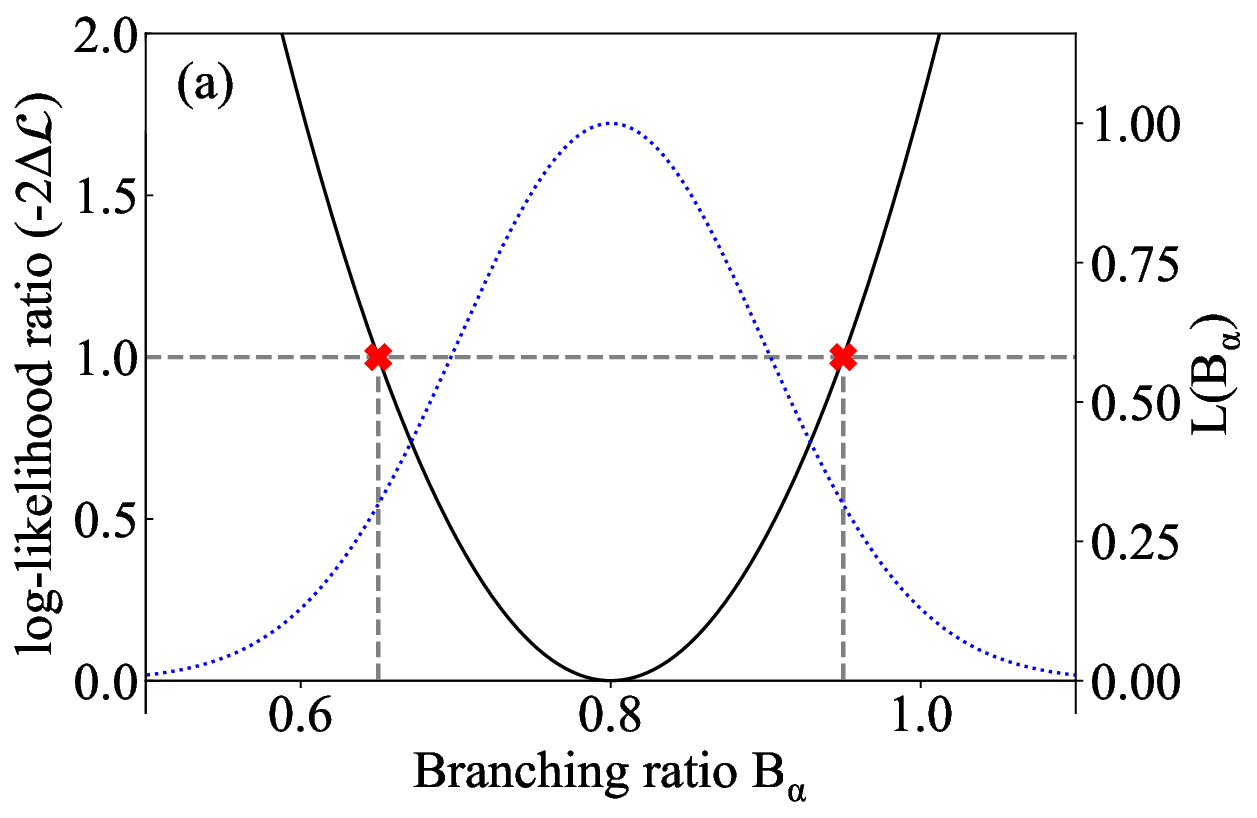}
\end{subfigure}

\begin{subfigure}
\centering
\includegraphics[width = 0.7\textwidth]{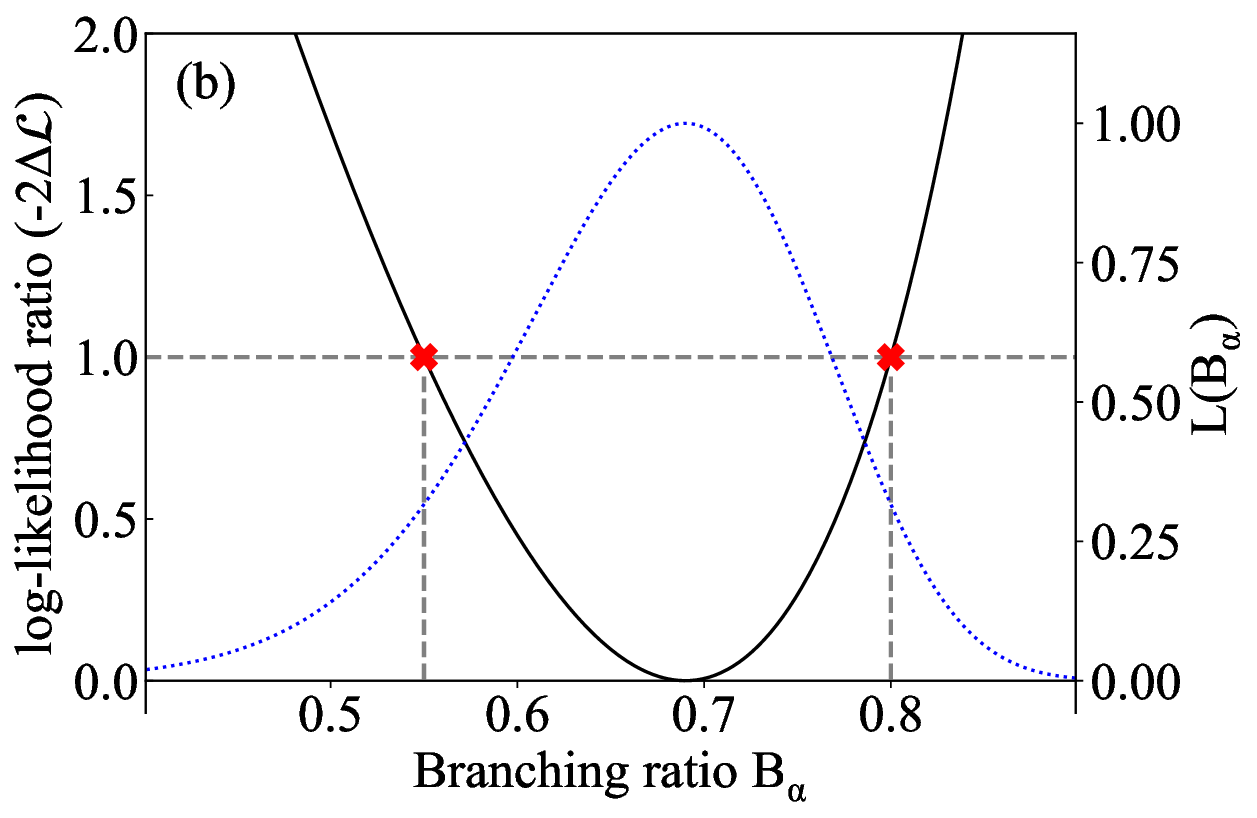}
\end{subfigure}
\caption{The log-likelihood ratio $-2\Delta\mathscr{L}$ curves (black solid) and likelihoods $L(B_\alpha)$ (blue dotted) reconstructed from the reported $\alpha-$branching ratios $B_\alpha$ in Refs.~\cite{2004Vi,2009Ta}. The branching ratio of E$_x$ = 4708.4~keV was measured as (a) $B_\alpha=0.80\pm0.15$ \cite{2009Ta}, and (b) $B_\alpha = 0.69 ^{+0.11}_{-0.14}$ \cite{2004Vi}. The red markers represent the one-sigma confidence levels, where the value of $-2\Delta\mathscr{L}$ becomes 1.}
\label{fig: log-likelihood_example_0}
\end{figure}

For asymmetric uncertainties expressed as ${\theta_0}^{+\sigma_1}_{-\sigma_2}$, similarly to the symmetric case, the log-likelihood ratio $-2\Delta\mathscr{L}(\theta)$ should satisfy conditions $-2\Delta\mathscr{L}(\theta_0+\sigma_1) = 1$ and $-2\Delta\mathscr{L}(\theta_0-\sigma_2) = 1$. Due to the asymmetric nature of the uncertainty, the quadratic approximation in Eq.~(\ref{eqn: Taylor}) does not adequately reconstruct the shape of the $-2\Delta\mathscr{L}(\theta)$ curve. Therefore, the revised quadratic forms adopted from Ref.~\cite{2004Ba} are used to appropriately reconstruct the asymmetric likelihood:

\begin{align}
\mathscr{L}(\theta)=-\frac{1}{2}\left( \frac{(\theta-\theta_0)}{\sigma+\sigma'(\theta-\theta_0)} \right)^2
\label{eqn: variable_1}
\end{align}

\begin{align}
\mathscr{L}(\theta)=-\frac{1}{2}\frac{(\theta-\theta_0)^2}{V+V'(\theta-\theta_0)},
\label{eqn: variable_2}
\end{align}

\noindent where $\sigma=\frac{2\sigma_1\sigma_2}{\sigma_1+\sigma_2},~\sigma'=\frac{\sigma_1-\sigma_2}{\sigma1+\sigma2},~V=\sigma_1\sigma_2$, and $V'=\sigma_1-\sigma_2$. In this work, the likelihood is modeled using an arithmetic combination of Eqs.~(\ref{eqn: variable_1}) and (\ref{eqn: variable_2}), where its effectiveness has been demonstrated in Ref.~\cite{2014Ra}. Fig.~\ref{fig: log-likelihood_example_0} (b) shows the reconstructed $-2\Delta \mathscr{L}(B_\alpha)$ curves and $L(B_\alpha)$ from the reported $\alpha-$branching ratios $B_\alpha$ in Ref.~\cite{2004Vi}, as an example.

\subsubsection{\label{sec: one-sided} Upper/Lower limit}

The reconstruction procedure becomes more complicated when only an upper or lower limit is reported. In this work, statistical assumptions and the approach proposed in Ref.~\cite{2011Co} are adopted. Suppose an experimental study reports that a model parameter $\theta$ satisfies $\theta < \theta_{\mathrm{up}}$ at the  90\% confidence level, with the MLE denoted as $\theta_0$. This situation corresponds to an extreme case of asymmetric uncertainty ${\theta_0}^{+\sigma_1}_{-\sigma_2}$ where $\sigma_2$ goes to infinity. In such a case, any value $\theta < \theta_0$ cannot be considered less or more probable than $\theta_0$, i.e., $t_{\theta<\theta_0}=t_{\theta_0}=0$. For the region $\theta \geq \theta_0$, the test statistic $t_\theta$ can be approximated by a quadratic function using the Taylor expansion, as shown in Eq.~(\ref{eqn: Taylor}). The complete form of $t_\theta$ is expressed as

\begin{align}
t_\theta=-2\Delta\mathscr{L}(\theta)=
\begin{cases}
0 & \text{for } \theta_0 \geq \theta \\
\dfrac{(\theta - \theta_0)^2}{2\sigma^2} & \text{for } \theta_0 < \theta,
\end{cases}
\label{eqn: half-gaussian}
\end{align}
\noindent where $\sigma$ is inferred from a statistical relationship between $\theta_0$ and $\theta_{up}$. The $p$-value in Eq.~(\ref{eqn: pvalue}) is computed using a half-chi-square distribution rather than a full chi-square distribution~\cite{2011Co}. At 90\% confidence level, the threshold value of $t_\theta$ is 1.64: $t_{\theta_{up}}=1.64=\dfrac{(\theta_{up}-\theta_0)^2}{2\sigma^2}$, or equivalently, $\theta_{up}=\theta_0+1.282\sigma$. This leads to the determination of $\sigma$, and thereby defines the corresponding half-Gaussian likelihood function.

\begin{figure}
\includegraphics[width=0.7\textwidth]{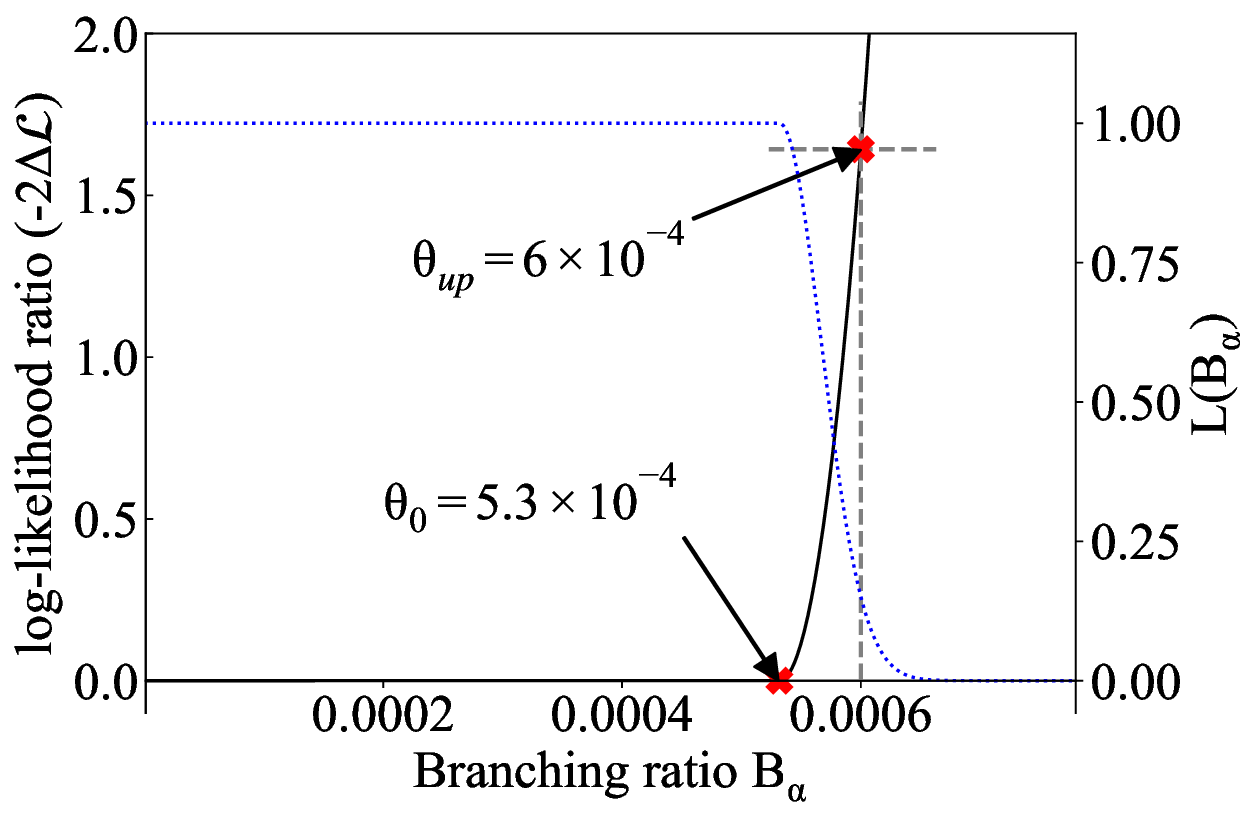}
\caption{The log-likelihood ratio $-2\Delta\mathscr{L}$ curve (black solid) and likelihood function $L(B_\alpha)$ (blue dotted) reconstructed from the $\alpha$-branching ratio measurement report $B_\alpha<6\times10^{-4}$ for E$_x$ = 4034~keV state~\cite{2003Re}. The red markers represent the MLE $\theta_0$ and the upper limit $\theta_{up}$, where $\theta_{up}=1.128\theta_0$ as determined by the statistical relation described in the text.}
\label{fig: log-likelihood_example_half}
\end{figure}

In practice, however, many experimental results report only the upper bound $\theta_{up}$ and the corresponding confidence level, while omitting the MLE $\theta_0$ and associated uncertainty $\sigma$. To reconstruct the likelihood under these conditions, we assume $\sigma = 0.1 \times \theta_0$. Substituting this into the above expression yields $\theta_{up}=\theta_0+1.282\sigma=1.128\theta_0$, from which $\theta_0$ can be inferred. This fully determines the half-Gaussian profile in Eq.~(\ref{eqn: half-gaussian}). Figure~\ref{fig: log-likelihood_example_half} shows an example of the reconstructed log-likelihood ratio curve $-2\Delta \mathscr{L}$ and likelihood function $L(B_\alpha)$ reconstructed from the previous report on $B_\alpha$ of E$_x$ = 4034~keV state~\cite{2003Re}.

\subsection{\label{sec:advantage}Implication}

This method is consistent with the conventional weighted averaging approach. Suppose a resonance parameter $\theta$ is initially given as $\theta_1 \pm \sigma_1$ and a new measurement reports $\theta = \theta_2 \pm \sigma_2$. If both the prior and likelihood follow Gaussian distributions, the posterior distribution $P(\theta\mid X)$ is also Gaussian:
\begin{align}
P(\theta\mid X) = \dfrac{\pi(\theta)L(\theta\mid X)}{\int\pi(\theta)L(\theta\mid X)d\theta} \propto e^{-\frac{(\theta-\theta_1)^2}{2\sigma_1^2}}\times e^{-\frac{(\theta-\theta_2)^2}{2\sigma_2^2}}.
\end{align}
\noindent By completing the square in the exponent and normalization, the posterior is identified as a normal distribution with mean $\theta_0 = \left(\frac{\theta_1}{\sigma_1^2}+\frac{\theta_2}{\sigma_2^2}\right)/\left(\frac{1}{\sigma_1^2}+\frac{1}{\sigma_2^2}\right)$ and standard deviation $\sigma_0 = 1/\sqrt{\left(\frac{1}{\sigma_1^2}+\frac{1}{\sigma_2^2}\right)}$. This is mathematically equivalent to those obtained through weighted averaging, demonstrating the consistency of the Bayesian approach with standard statistical methods under the assumption of Gaussian distributions.

An additional advantage of this method lies in its rigorous treatment of asymmetric uncertainties and upper/lower limits—features that are inherently challenging to address the weighted averaging frameworks. This capability becomes particularly valuable when dealing with inconsistent measurement and results presented in non-Gaussian forms.

\begin{figure}
\includegraphics[width=0.7\textwidth]{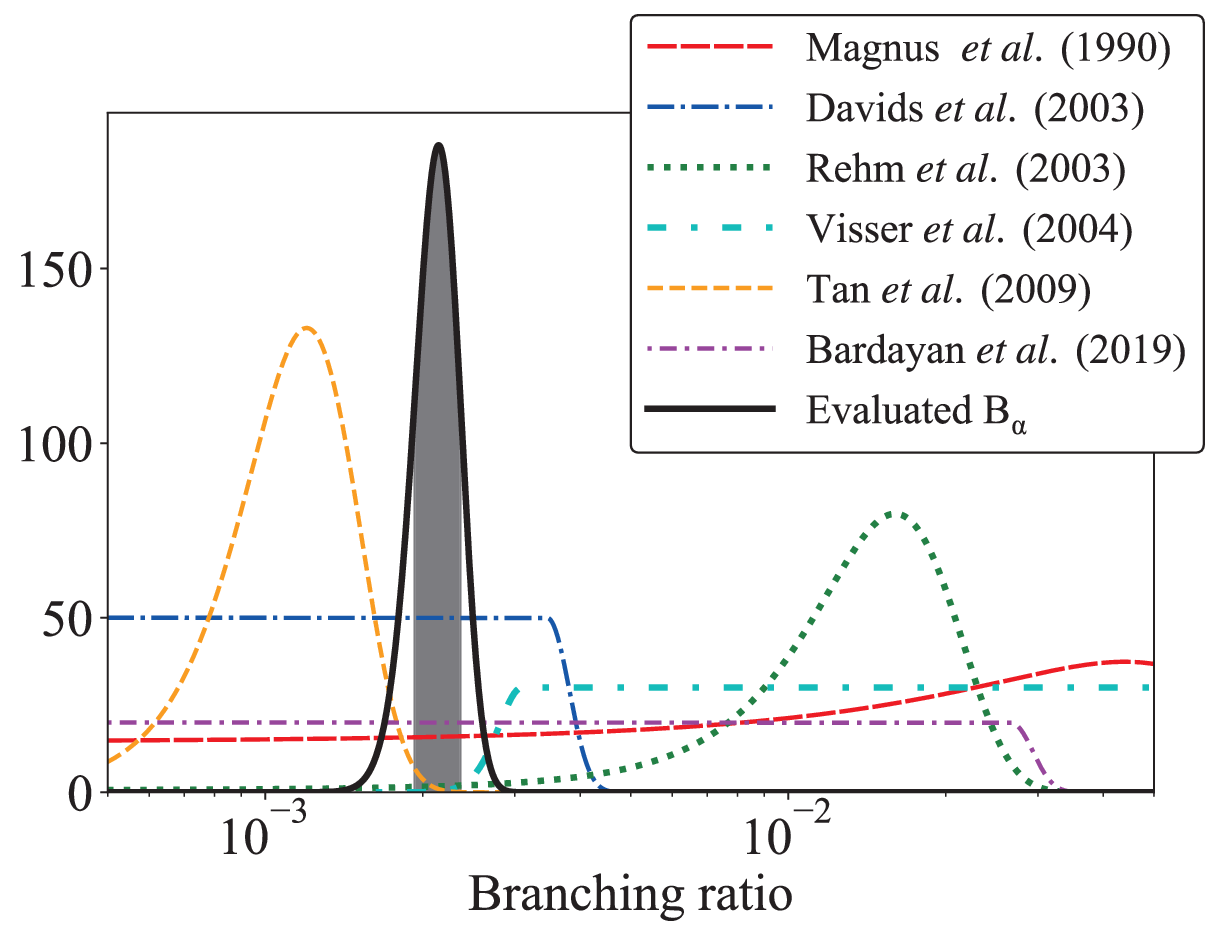}
\caption{Individual likelihood functions $L(B_\alpha)$ derived from the $B_\alpha$ measurements for E$_x$ = 4377~keV summarized in Table~\ref{tab:Ba measurements} are shown as colored-dashed lines. The final evaluated posterior distribution is represented by the black solid line, with the shaded band indicating the 1-$\sigma$ credible interval.}
\label{fig: Ba measurements}
\end{figure}

A representative example is the set of $B_\alpha$ measurements for the E$_x = $ 4377~keV state, which exhibit significant discrepancies between measurements \cite{1990Ma,2003Da,2003Re,2004Vi,2009Ta,2019Ba} as summarized in Table~\ref{tab:Ba measurements}. Previous evaluations of the $^{15}$O($\alpha$, $\gamma$)$^{19}$Ne reaction rate have selectively adopted only a subset of these data~\cite{2011Da,2009Ta}, as a statistically rigorous combination was not straightforward. In contrast, the Bayesian framework facilitates a consistent and coherent incorporation of all available data. The reconstructed likelihood distributions shown in Fig.~\ref{fig: Ba measurements} allow a robust estimation of the posterior distribution and its associated uncertainty, even in the presence of asymmetric errors or one-sided limits.

\begin{table}
\begin{ruledtabular}
    
    \renewcommand{\arraystretch}{1.5}
    \caption{B$_a$ of E$_x$ = 4377~keV state measurements}
    \begin{tabular}{ll}
         Ref.& $B_\alpha$\\
         \hline
         Magnus \textit{et al.} (1990) \cite{1990Ma}& $4.4\pm3.2\times10^{-2}$  \\
         Davids \textit{et al.} (2003) \cite{2003Da}& $<3.9\times10^{-3}$ \\
         Rehm \textit{et al.} (2003) \cite{2003Re}& $1.6\pm0.5\times10^{-2}$ \\
         Visser \textit{et al.} (2004) \cite{2004Vi}& $>2.7\times10^{-3}$ \\
         Tan \textit{et al.} (2009) \cite{2009Ta}& $1.2\pm0.3\times10^{-3}$ \\
         Bardayan \textit{et al.} (2019) \cite{2019Ba}& $<3.0\times10^{-2}$\\ 
    \end{tabular}
   
    \label{tab:Ba measurements}
\end{ruledtabular}

\end{table}

\section{\label{sec:result}$^{19}$Ne Evaluation}
\subsection{\label{sec:spin}Spin-parity}

In most cases, experimental reports of spin-parity assignments $J^\pi$ provide only a single most probable value, without an accompanying estimate of uncertainty. This absence of uncertainty information presents a challenge for their incorporation into a Bayesian framework, which requires well-defined probabilistic likelihoods. To address this limitation, a surrogate method inspired by the approach of Mohr $\textit{et al.}$~\cite{2014Mo} is employed to construct the probability distribution $p(J^\pi)$ over the possible assignment.

When experimental studies report multiple, conflicting $J^\pi$ assignments for a given excited state, a uniform probability distribution is assumed over all reported values. However, if a degree of consensus is apparent—i.e., one assignment is more frequently reported or supported by stronger experimental evidence—that assignment is given a dominant prior weight of 50\%, and the remaining 50\% distributed uniformly among the alternative possibilities.

Additionally, mirror symmetry considerations are incorporated to refine the distribution $p(J^\pi)$. Specifically, if the $J^\pi$ of the mirror level in the analog nucleus $^{19}$F is known with confidence, the corresponding assignment in $^{19}$Ne is given increased weight. This consideration is particularly valuable for states where experimental ambiguities persist.

Furthermore, mirror symmetry is utilized to constrain $p(J^\pi)$ by assigning enhanced prior weight to spin-parity values corresponding to well-established mirror levels in the analog nucleus $^{19}$F. This provides an additional layer of inference in cases where direct experimental assignments in $^{19}$Ne are uncertain or conflicting. Table~\ref{tab: pJ} summarizes the spin-parity assignments for all relevant excited states. The corresponding $p(J^\pi)$ are explained in Sec.~\ref{sec: level information}.

{\scriptsize
\renewcommand{\arraystretch}{1.3}
\begin{longtable}[l]{llllll}
\caption{Spin-parity assignments for excited states in $^{19}$Ne. For levels with multiple possible assignments, the recommended one is indicated with an asterisk (*), corresponding to the highest value of $p(J^\pi)$ in this work. Assignments with equal $p(J^\pi)$ values are grouped in a single row, while those with different probabilities $p(J^\pi)$ are listed in separate lines (e.g., for E$_x$ = 7995~keV, $p(\frac{5}{2}^+)=0.5$ and remaining $p(J^\pi)$ is distributed equally to $J^\pi=\frac{1}{2}^+$ and $\frac{13}{2}^+$ as 0.25.). The values of $p(J^\pi)$ are explained in Sec.~\ref{sec: level information} if needed.\label{tab: pJ}}\\

\hline\hline
E$_x$ [keV] &$J^{\pi}$  &Ref &E$_x$ [keV] &$J^{\pi}$  &Ref\\\cmidrule{1-3} \cmidrule{4-6}
\endfirsthead

\hline
E$_x$ [keV] &$J^{\pi}$  &Ref &E$_x$ [keV] &$J^{\pi}$  &Ref\\\cmidrule{1-3} \cmidrule{4-6}

\endhead

\multicolumn{6}{c}{\textit{(Continued on next page)}}\\
\hline
\endfoot

\hline\hline
\endlastfoot

4034 &$\dfrac{3}{2}^+$ &\cite{1970Ga,1971Bi,1973Da,2015Pa} &6435 &${\dfrac{1}{2}^-}^*$ &\cite{2007Ne,2017To,2022Go} \\
4143 &$\dfrac{7}{2}^-$ &\cite{1971Bi,1973Da,2005Ta,2009Ta,2015Pa} & &$\dfrac{11}{2}^+$ &\cite{2013La} \\
4200 &$\dfrac{9}{2}^-$ &\cite{1971Bi,1973Da,2005Ta,2009Ta,2015Pa} &6450 & ${\dfrac{3}{2}^+}^*$ &\cite{2020Ha,2015Ch} \\
4377 &$\dfrac{7}{2}^+$ &\cite{1974Ga,1973Da,1971Bi} & &$\dfrac{5}{2}^-$ &\cite{2013La} \\
4548 &$\dfrac{3}{2}^-$ &\cite{1970Ga,1971Bi,1973Da,2015Pa} &6537 &$\dfrac{7}{2}^+$ &\cite{2022Go,2017To,2015Ch} \\
4602 &$\dfrac{5}{2}^+$ &\cite{2015Pa,1973Da} &6700 &$\dfrac{5}{2}^-, \dfrac{5}{2}^+$ &\cite{2022Go,2005Ko} \\
4634 &$\dfrac{13}{2}^+$ &\cite{1971Bi,1972Pa,1973Da} &6743 &$\dfrac{3}{2}^-$ &\cite{2022Go,2011Ad(b),2013La,2015Ba,2015Ch} \\
4708 &$\dfrac{5}{2}^-$ &\cite{2015Pa} &6851 &${\dfrac{5}{2}^+}^*$ &\cite{1995NDS,1972Pa} \\
5091 &$\dfrac{5}{2}^+$ &\cite{1971Bi,2015Ba} & &$\dfrac{9}{2}^-, \dfrac{11}{2}^-, \dfrac{11}{2}^+, \dfrac{13}{2}^+$ & \\
5351 &$\dfrac{1}{2}^+$ &\cite{2017To,2017Ka,1971Bi,1970Ga,2022Go} &6863 &$\dfrac{7}{2}^-$ &\cite{2013La,2023Po} \\
5424 &$\dfrac{7}{2}^+$ &\cite{1971Bi,1972Pa} &6967 &$\dfrac{5}{2}^+$ &\cite{2022Go,2015Ch} \\
5488 &$\dfrac{3}{2}^+$ &\cite{2017Ka,2017To,2022Go} &7027 &$\dfrac{1}{2}^-$ &\cite{2022Go} \\
5535 &$\dfrac{7}{2}^-, \dfrac{5}{2}^+$ &\cite{1995NDS} &7072 &$\dfrac{3}{2}^+$ &\cite{2011Ad(b),2015Ch,2015Ba,2015Pa,2017Ka,2012Mo,2009Da,2009Mu} \\
5704 &$\dfrac{5}{2}^-$ &\cite{2017To} &7173 &$\dfrac{11}{2}^-$ &\cite{2015Pa} \\
5830 &${\dfrac{1}{2}^+}^*$ &\cite{2017Ka,1995NDS} &7230 &$\dfrac{3}{2}^+$ &\cite{2012Mo,2004Ba_19Ne} \\
&$\dfrac{3}{2}^+$ & &7279 &${\dfrac{1}{2}^+}^*$ &\cite{2022Go,2004Ba_19Ne} \\
6014 &$\dfrac{3}{2}^-$ &\cite{2022Go,2017To,2013La,1972Ha,1970Ga} & &$\dfrac{3}{2}^+$ &\cite{2009Da} \\
6081 &$\dfrac{5}{2}^-, \dfrac{3}{2}^+$ &\cite{2013La} &7396 &$\dfrac{7}{2}^+$ &\cite{2022Go,2017To,2004Ba_19Ne} \\
6100 &$\dfrac{7}{2}^+$ &\cite{2013La,2020Ha} &7495 &$\dfrac{5}{2}^+$ &\cite{2009Da,2009Mu,2017To,2022Go,2012Mo} \\
6138 &$\dfrac{1}{2}^+, \dfrac{3}{2}^+$ &\cite{2013La,2017To,2017Ka,2022Go,2023Po} &7532 &$\dfrac{5}{2}^-$ &\cite{2004Ba_19Ne} \\
6272 &$\dfrac{7}{2}^+$ &\cite{2022Go} &7615 &$\dfrac{3}{2}^+$ &\cite{2009Da,2017To,2022Go,2012Mo} \\
6279 &$\dfrac{5}{2}^+$ &\cite{2017To,2022Go} &7668 &$\dfrac{3}{2}^-$ &\cite{2009Mu,2022Go,2012Mo} \\
6286 &$\dfrac{1}{2}^+, \dfrac{3}{2}^+$ &\cite{2011Ad(b),2013La,2015Ba} &7764 &${\dfrac{3}{2}^+}^*$ &\cite{2012Mo,2009Mu} \\
6292 &$\dfrac{11}{2}^+$ &\cite{2020Ha} & &$\dfrac{1}{2}^+$ &\cite{2017Ka} \\
6417 &$\dfrac{3}{2}^-$ &\cite{2013La,2011Ad} &7872 &$\dfrac{1}{2}^+$ &\cite{2022Go,2017To,2012Mo,2009Da} \\
6423 &$\dfrac{3}{2}^+$ &\cite{2017Ka,2020Ha} &7995 &${\dfrac{5}{2}^+}^*$ &\cite{2012Mo} \\
& & & &$\dfrac{1}{2}^+, \dfrac{5}{2}^-$ &\cite{2009Da,2022Go,2009Mu} \\


\end{longtable}
}

\subsection{\label{sec: mirror} Mirror analysis}

When resonance parameters for a certain energy level in $^{19}$Ne are sparsely measured or entirely unavailable, mirror symmetry in invoked to estimate the corresponding quantities. Analog states in the mirror nucleus $^{19}$F~\cite{1995NDS,2017To,2022Go} are used as proxies for the corresponding levels in $^{19}$Ne, following an approach established in prior studies~\cite{2007Ne} and applied here within a unified framework.

For each resonance level in \(^{19}\mathrm{Ne}\), the reduced partial width \(\theta_{\lambda}^2\) is assumed to be identical to that of its mirror counterpart in \(^{19}\mathrm{F}\), based on the assumption of charge symmetry~\cite{2007Ne}. The corresponding resonance widths in \(^{19}\mathrm{Ne}\) are derived by scaling the measured widths in \(^{19}\mathrm{F}\) using appropriate kinematic and penetrability factors. For example, the \(\alpha\)-decay width \(\Gamma_{\alpha}\) in \(^{19}\mathrm{Ne}\) is calculated as:
\begin{align}
    \Gamma_{\alpha,\,^{19}\mathrm{Ne}} &= \left[ \frac{2\hbar P_{\ell}}{\mu R^2} \right]_{\mathrm{^{15}O}+\alpha} \theta_{\alpha,\,^{19}\mathrm{F}}^2 \label{eq:gamma_alpha} \\
    &= \left( \frac{P_{\ell}}{\mu} \right)_{\mathrm{^{15}O}+\alpha} 
       \left( \frac{\mu}{P_{\ell}} \right)_{\mathrm{^{15}N}+\alpha} 
       \Gamma_{\alpha,\,^{19}\mathrm{F}}. \label{eq:gamma_alpha_mirror}
\end{align}

A similar approach is applied for proton decay widths:
\begin{align}
    \Gamma_{p,\,^{19}\mathrm{Ne}} &= \left[ \frac{2\hbar P_{\ell}}{\mu R^2} \right]_{\mathrm{^{18}F}+p} \theta_{p,\,^{19}\mathrm{F}}^2 \notag \\
    &= \left( \frac{P_{\ell}}{\mu} \right)_{\mathrm{^{18}F}+p}
       \left( \frac{\mu}{P_{\ell}} \right)_{\mathrm{^{18}O}+p}
       \Gamma_{p,\,^{19}\mathrm{F}}.
\end{align}

This same formalism is applied to the total width \(\Gamma\), which is often inferred from lifetime measurements near the \(\alpha\)-threshold. In cases where $\Gamma$ is dominated by $\Gamma_\alpha$--as is typically the case for higher-lying resonances--the use of mirror symmetry provides a valid approximation. Notably, the \(\alpha\)-branching ratio \(B_{\alpha} = \Gamma_{\alpha}/\Gamma\) is preserved under mirror symmetry, since it is a dimensionless observable that characterizes the intrinsic decay properties of the state and is independent of absolute width scaling.

The mirror analysis presented in this work is restricted to levels below E\(_x = 6014\)~keV, where the analog assignments are relatively well established. At higher excitation energies, the lack of definitive mirror-state identifications introduces significant uncertainties, and the method is therefore not extended to those states.

\subsection{\label{sec: tau_gamma}Mean lifetime $\tau_m$ to decay width $\Gamma$ conversion}

\begin{figure}
\begin{subfigure}
\centering
\includegraphics[width=0.45\textwidth]{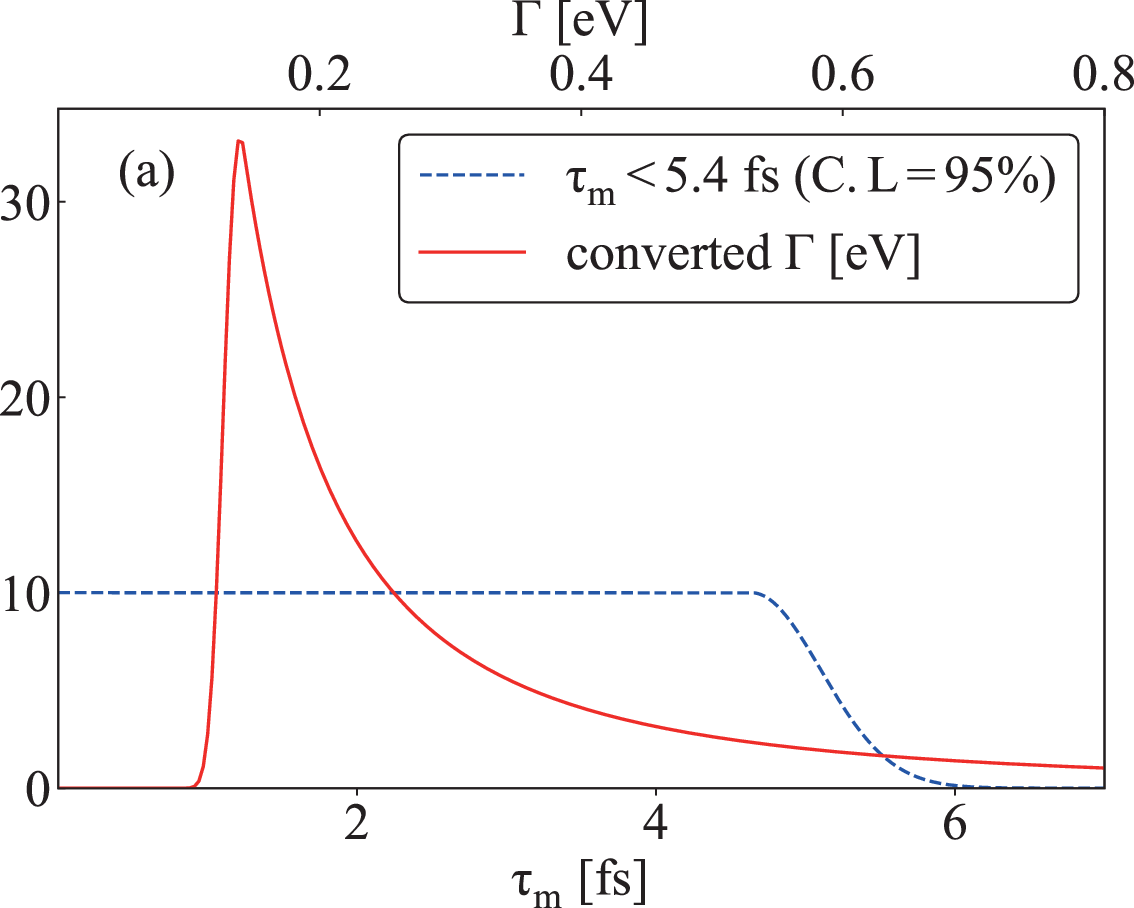}
\end{subfigure}
\begin{subfigure}
\centering
\includegraphics[width=0.45\textwidth]{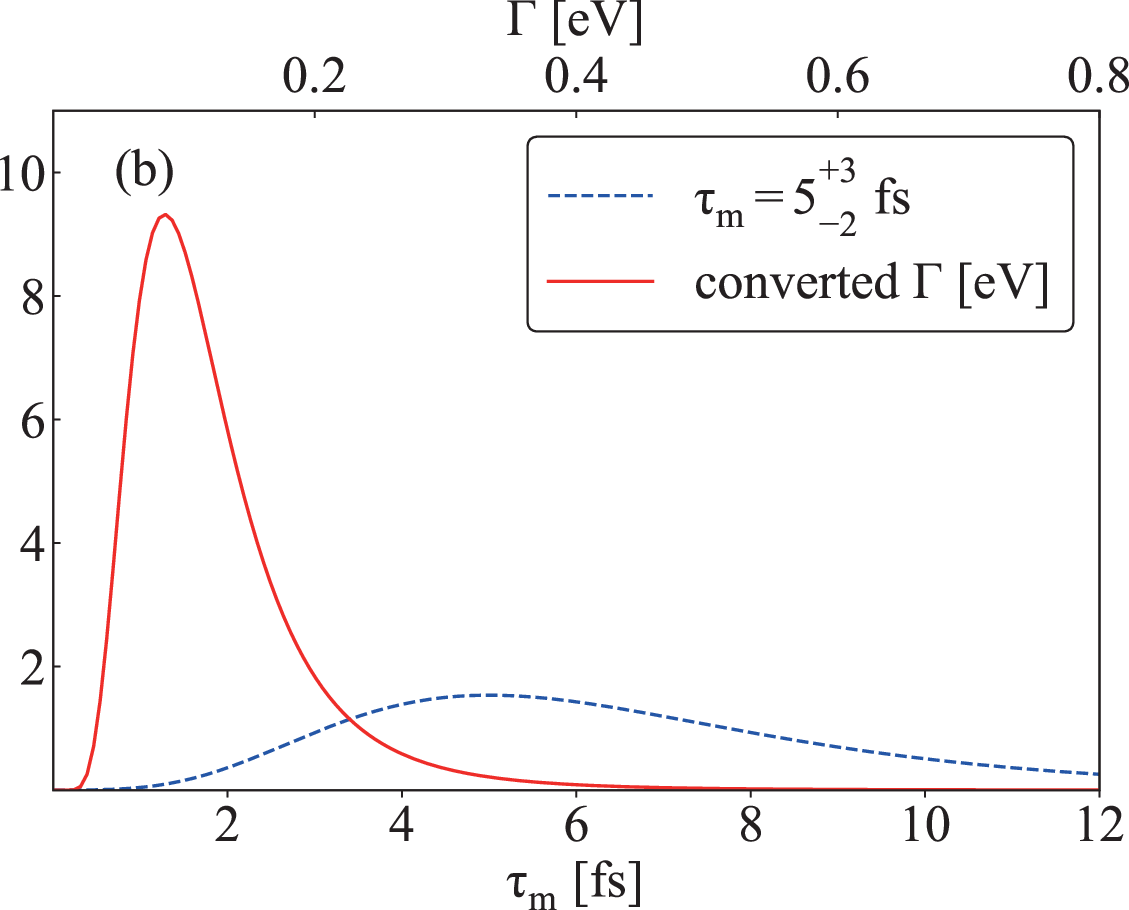}
\end{subfigure}
\caption{Examples of converting likelihood functions for the mean lifetime $\tau_m$ (blue dashed) into those for the decay width $\Gamma$ (red solid). The likelihood functions in (a) and (b) are reconstructed from the measurement reports of $\tau_m<5.4$~fs~\cite{2008My} and $\tau_m=5^{+3}_{-2}$~fs~\cite{2005Ta} (both for E$_x$ = 4377~keV state), respectively.}
\label{fig:tau_to_gamma}
\end{figure}

\begin{figure}
\includegraphics[width=0.7\textwidth]{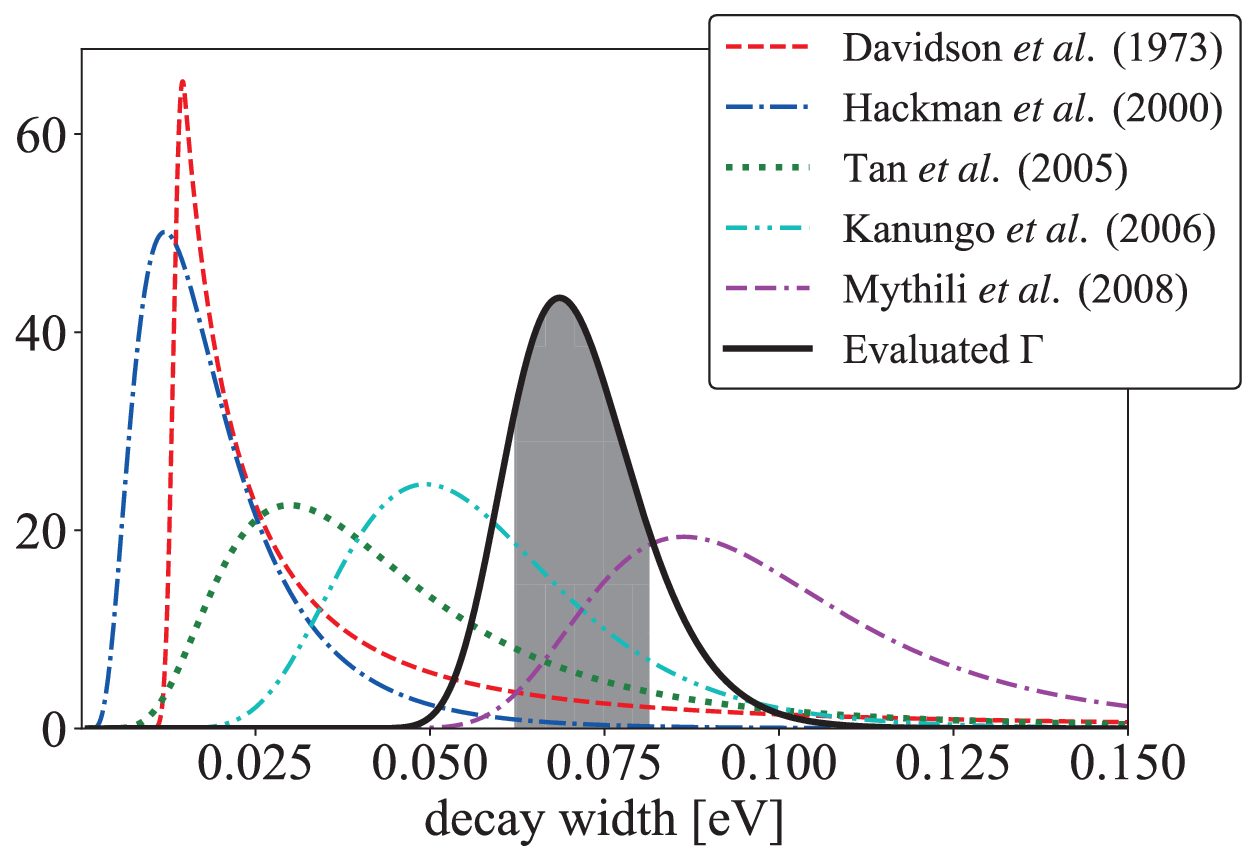}
\caption{Individual likelihood functions $L(\Gamma)$ for the decay width of the E$_x$ = 4034~keV state, derived from various experimental measurements (colored dashed lines). The evaluated posterior distribution is shown as a solid black line, with the shaded region indicating the 1$\sigma$ credible interval.}
\label{fig: Ex4034 total width}
\end{figure}

In several cases, particularly for states near the \(\alpha\)-threshold, the total decay width \(\Gamma\) of a resonance has been inferred from experimental measurements of the mean lifetime \(\tau_m\). To incorporate this information into the evaluation framework, the likelihood distribution \(L(\tau_m)\), derived from lifetime data, is transformed into the corresponding likelihood for \(\Gamma\) using the relation,  $\Gamma = \frac{\hbar}{\tau_m}$.

The transformation of the likelihood function follows the standard rule for change of variables in probability theory. Specifically, the likelihood for \(\Gamma\) is expressed as:
\begin{equation}
    L(\Gamma) = L(\tau_m) \left| \frac{\partial\tau_m}{\partial\Gamma} \right| 
    = L\left( \frac{\hbar}{\Gamma} \right) \frac{\hbar}{\Gamma^2}. \label{eq:lifetime_to_width}
\end{equation}

\noindent Typical examples of transformed likelihood distributions are shown in Fig.~\ref{fig:tau_to_gamma}, illustrating how a half normal or variable Gaussian distribution (Eqs.~(\ref{eqn: variable_1}) and (\ref{eqn: variable_2})) in \(\tau_m\) space is mapped onto a distribution in \(\Gamma\) space.

To ensure consistency, this transformation is systematically applied wherever $\tau_m$ data are available. As an example, for the E$_x=4034$~keV level,  four independent lifetime-based likelihoods are transformed and combined with a direct $\Gamma$ measurement~\cite{1973Da, 2005Ta, 2006Ka, 2008My, 2000Ha}. The resulting posterior distribution (Fig.~\ref{fig: Ex4034 total width}) yields a statistically well-constrained estimate of the total width.

\subsection{\label{sec: level information} Level information}

The nuclear structure properties of 46 excited states in \(^{19}\mathrm{Ne}\), with excitation energies ranging from 4034 to 7995~keV, are evaluated with the Bayesian approach detailed above, incorporating all experimental data reported between 1967 and 2022. For each resonance parameter, the median of the posterior probability distribution is adopted as the representative value. The associated uncertainty range is defined by the 16th and 84th percentiles, corresponding to a one-sigma credible interval. In cases where the one-sigma interval does not adequately reflect the shape of the posterior distribution, such as for highly skewed or truncated ones, the 90th percentile is given to better characterize the upper or lower bound. Detailed evaluations and justifications are provided in the following subsections.

\begin{figure}
    \centering
    \includegraphics[width=0.8\linewidth]{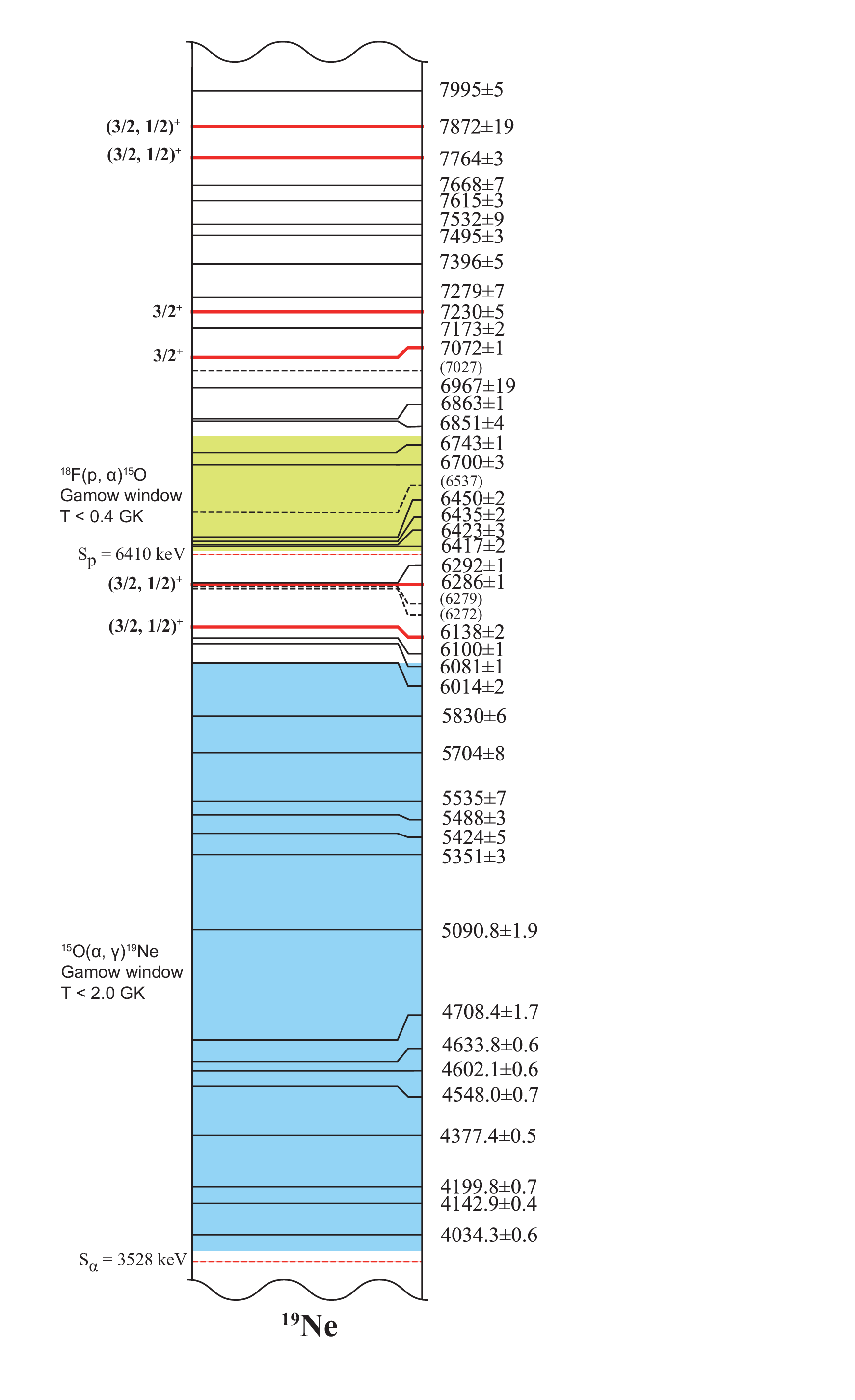}
    \caption{Level scheme of $^{19}$Ne. The unit of excitation energy is keV. The blue- and green-shaded regions represent the Gamow windows for the $^{15}$O($\alpha, \gamma$)$^{19}$Ne and the $^{18}$F($p, \alpha$)$^{15}$O reactions, respectively. The important levels out of the Gamow window are marked by red with their spin-parity $J^\pi$ assignments.}
    \label{fig:level scheme}
\end{figure}

To enhance clarity and astrophysical relevance, the evaluation is organized into three energy regions. The first, from E\(_x = 4034~\mathrm{keV}\) (S$_\alpha$) to E\(_x = 6014~\mathrm{keV}\), comprises levels that contribute to the \(^{15}\mathrm{O}(\alpha,\gamma)^{19}\mathrm{Ne}\) reaction in XRBs. The second, from E\(_x = 6014~\mathrm{keV}\) to E\(_x = 6292~\mathrm{keV}\), encompasses potential subthreshold resonances relevant to the \(^{18}\mathrm{F}(p,\alpha)^{15}\mathrm{O}\) reaction in nova at low temperatures (\(T_9 < 0.3\)~GK). The third region covers states above E\(_x = 6417~\mathrm{keV}\) (S$_p$), which serve as resonances in the \(^{18}\mathrm{F}(p,\alpha)^{15}\mathrm{O}\) reaction. Figure~\ref{fig:level scheme} shows a level scheme of $^{19}$Ne.

When available, mirror-state information in \(^{19}\mathrm{F}\) is used to constrain uncertain parameters in $^{19}$Ne. Mirror assignments are primarily based on the Ref.~\cite{1995NDS}, supplemented by more recent structural information from Refs.~\cite{2011Co, 2005Ba}.

\subsubsection{From E$_x$ = 4034~keV (S$_\alpha$) to E$_x$ = 6014~keV}
The resonance parameters required for calculation of the $^{15}$O($\alpha,\gamma$)$^{19}$Ne reaction rates in stellar temperatures and the relevant references are summarized in Table~\ref{tab:Tab2}. For levels in this section, the corresponding mirror states are obvious. Therefore, the unmeasured parameters required to calculate the reaction rates are adopted directly from their mirror counterparts.

\begin{turnpage}
    
\begin{table*}
\begin{ruledtabular}
\caption{Evaluated resonance parameters in this work for energy levels in $^{19}$Ne that contribute to the $^{15}$O($\alpha, \gamma$)$^{19}$Ne reaction at stellar temperatures ($<2$~GK). Asterisks indicate values adopted from the mirror nucleus. The evaluated spin-parity assignments are presented. The references for spin-parity are in Table~\ref{tab: pJ}.}\label{tab:Tab2}
\footnotesize
\begin{tabular}{ccccccccc}
\multirow{2}{*}{E$_x$ [keV]} &\multirow{2}{*}{$J^\pi$} &\multirow{2}{*}{$B_\alpha$} &\multirow{2}{*}{$\Gamma$ [eV]} &\multirow{2}{*}{$\Gamma_\alpha$ [keV]} &\multicolumn{4}{c}{Reference} \\\cmidrule{6-9}
& & & & &E$_x$ &$B_\alpha$ &$\Gamma$ &$\Gamma_\alpha$ \\
\hline
4034.3$\pm$0.6& $\frac{3}{2}^+$&$2.4^{+1.2}_{-1.4}\times10^{-4}$&$7.1^{+1.0}_{-0.9}\times10^{-2}$& &\cite{1967Gr,1968Gu,1970Ga,1971Bi,1973Da,2005Ta,2015Do,2015Ba,2020Ha}&\cite{2002La,2003Da,2003Re,2009Ta} &\cite{1973Da,2000Ha,2005Ta,2006Ka,2008My} & \\
4142.9$\pm$0.4&$\frac{7}{2}^-$ &$(1.2\pm0.5)\times10^{-3}$ &$4.1^{+0.8}_{-0.6}\times10^{-2}$& &\cite{1967Gr,1968Gu,1970Ga,1971Bi,1973Da,2005Ta,2015Do,2015Ba,2019Ha}&\cite{2002La,2009Ta} &\cite{1973Da,2005Ta,2008My} & \\
4199.8$\pm$0.7& $\frac{9}{2}^-$&$(1.2\pm0.5)\times10^{-3}$ &$(1.5\pm{0.3})\times10^{-2}$& &\cite{2005Ta,2015Ba,2015Do,2019Ha,1970Ga,1971Bi,1973Da,1968Gu}&\cite{2002La,2009Ta} &\cite{1973Da,2005Ta,2008My} & \\
4377.4$\pm$0.5& $\frac{7}{2}^+$&$(2.1\pm0.2)\times10^{-3}$&$0.16^{+0.07}_{-0.03}$& &\cite{1968Gu,1970Ga,1971Bi,1973Da,1974Ga,2005Ta,2015Ba,2015Pa,2020Ha}&\cite{1990Ma,2003Da,2003Re,2019Ba,2009Ta,2004Vi}&\cite{1973Da,2005Ta,2008My} & \\
4548.0$\pm$0.7& $\frac{3}{2}^-$&$0.08\pm0.01$ &$3.5^{+0.8}_{-0.6}\times10^{-2}$& &\cite{1968Gu,1998Ut,1970Ga,1971Bi,1973Da,2005Ta,2015Ba,2020Ha}&\cite{1990Ma,2003Da,2009Ta,2004Vi} &\cite{1973Da,2005Ta,2008My} & \\
4602.1$\pm$0.6& $\frac{5}{2}^+$&$0.26\pm0.01$ &$8.7^{+3.0}_{-1.9}\times10^{-2}$ & &\cite{1998Ut,1971Bi,1973Da,2005Ta,2020Ha}&\cite{1990Ma,2002La,2003Da,2009Ta,2004Vi} &\cite{1973Da,2005Ta,2008My} & \\
4633.8$\pm$0.6 & $\frac{13}{2}^+$&$<0.04^*$\footnotemark[1] &$<1.05\times10^{-3}$& &\cite{1968Gu,1970Ga,1971Bi,1972Pa,1973Da,2005Ta,2020Ha,1979Ma}&\cite{1995NDS} &\cite{1973Da,2005Ta} & \\
4708.4$\pm$1.7& $\frac{5}{2}^-$&$0.82\pm0.03$&$3.3^{+0.8}_{-0.5}\times10^{-2 *}$& &\cite{1968Gu,1970Ga,1971Bi,2020Ha}&\cite{1990Ma,2003Da,2009Ta,2004Vi} &\cite{1995NDS} & \\
5090.8$\pm$1.9& $\frac{5}{2}^+$&$0.82\pm0.02$ &$<0.11^*$& &\cite{1968Gu,1970Ga,1971Bi,1972Ha,1973Da,2015Ba,2020Ha,2011Ad(b)}&\cite{1990Ma,2002La,2003Da,2003Re,2009Ta,2019Ba,2004Vi} &\cite{1995NDS} & \\
5351$\pm$3 &$\frac{1}{2}^+$&$>0.66$&&$4.4\pm1.5$ &\cite{2017To,2017Ka,1971Bi,1970Ga,2011Ad(b)}&\cite{2002La} &&\cite{2017To,2006Va} \\
5424$\pm$5 & $\frac{7}{2}^+$&$>0.66$&$<8.9$& &\cite{1979Ma,1998Ut,1970Ga,1971Bi,1972Ha,1972Pa,2015Ba} &\cite{2002La} &\cite{1995NDS} & \\
5488$\pm$3& $\frac{3}{2}^+$&$>0.66$& &$9\pm2$ &\cite{1970Ga,2017Ka,2017To,2011Ad(b)}&\cite{2002La} & &\cite{2017To} \\
5535$\pm$7& $\frac{7}{2}^-/\frac{5}{2}^+$&$0.88\pm0.05$& &$2.1^{+0.6}_{-0.5}\times10^{-3 *}$ &\cite{1970Ga,1972Ha,2015Ba} &\cite{2019Ba} & &\cite{1995NDS} \\
5704$\pm$8& $\frac{5}{2}^-$& &$0.8^{+1.2 *}_{-0.3}$&$27\pm6$ &\cite{2017To} & &\cite{1995NDS} &\cite{2017To} \\
5830$\pm$6& $\frac{1}{2}^+$& & &$0.66\pm0.07^*$ &\cite{1970Ga,1972Ha,2015Ch,2017Ka} & & &\cite{1995NDS} \\
\end{tabular}
\footnotemark[1]{adopted from~\cite{1995NDS}}
\end{ruledtabular}
\end{table*}
\end{turnpage}

\textbf{1. E$_x$ = 4034.3$\pm$0.6 keV, $J^\pi$ = $\dfrac{3}{2}^+$}

This level is one of the most critical resonance levels contributing to the $^{15}$O($\alpha, \gamma$)$^{19}$Ne reaction rate. The spin-parity is consistently assigned as $J^\pi=\frac{3}{2}^+$, supported by multiple experimental studies~\cite{1970Ga,1971Bi,1973Da,2015Pa}. The two key resonance parameters for this state are the $\alpha$-branching ratio $B_\alpha$ and the total width $\Gamma$. The total width is inferred from five independent measurements~\cite{1973Da,2000Ha,2005Ta,2006Ka,2008My}, yielding an evaluated value of $0.071^{+0.010}_{-0.009}$ eV. Notably, the reported lifetime has decreased over time, from $\tau_m <$ 50~fs in 1973 to 6.9$\pm1.5\pm0.7$~fs in 2008 \cite{1973Da,2000Ha,2005Ta,2006Ka,2008My}, likely reflecting improvements in experimental resolution and detection sensitivity. This trend implies a larger decay width than previously estimated. The branching ratio $B_\alpha$ has been measured in four independent experiments~\cite{2002La,2003Da,2003Re,2009Ta}, and the evaluated $B_\alpha$ value in this level is $2.4^{+1.2}_{-1.4}\times10^{-4}$.

\textbf{2-3. E$_x$ = 4142.9$\pm$0.4 keV \& 4199.8$\pm$0.7 keV, $J^\pi$ = $\dfrac{7}{2}^-$ \& $\dfrac{9}{2}^-$ }

The two subsequent levels above the \(\alpha\)-threshold, at \(E_x = 4142.9 \pm 0.4~\mathrm{keV}\) and \(E_x = 4199.8 \pm 0.7~\mathrm{keV}\), have uncertain spin-parity assignments. Mirror-state comparison with \(^{19}\mathrm{F}\) suggest that the corresponding levels at \(E_x = 3998.7\) and \(4032.5~\mathrm{keV}\) levels in $^{19}$F have \(J^{\pi} = \frac{7}{2}^{-}\) and \(\frac{9}{2}^{-}\), respectively. Due to limited experimental resolution, unambiguous identification of the $^{19}$Ne levels with the $^{19}$F analogs has remained challenging.

Several interpretations of the spin-parity assignments have been proposed. Davidson~\textit{et al.}~\cite{1973Da} assigned the lower-energy \(^{19}\mathrm{Ne}\) state at \(4142.9~\mathrm{keV}\) as \(J^{\pi} = \frac{9}{2}^{-}\), based on the absence of expected \(\gamma\)-transitions. Parikh~\textit{et al.}~\cite{2015Pa} performed a DWBA analysis of angular distributions from the \(^{19}\mathrm{F}(^3\mathrm{He},t)^{19}\mathrm{Ne}\) reaction and concluded the same: \(J^{\pi} = \frac{9}{2}^{-}\) to the lower-energy state and \(J^{\pi} = \frac{7}{2}^{-}\) for the higher one. Subsequently, Tan~\textit{et al.}~\cite{2005Ta,2009Ta} proposed a reversal of the assignments, based on measured lifetimes and comparison with those of mirror states: \(18^{+2}_{-3}~\mathrm{fs}\) and \(43^{+12}_{-9}~\mathrm{fs}\) for the \(4142.8\) and \(4199.8~\mathrm{keV}\) levels, respectively. More recently, Hall~\textit{et al.}~\cite{2019Ha} observed a \(\gamma\)-transition from the 4142.9~keV state to a \(J^{\pi} = \frac{3}{2}^{-}\) level, further supporting the assignment of \(J^{\pi} = \frac{7}{2}^{-}\) for the 4142.9~keV state. Taking into account the full body of experimental evidence, this work adopts \(J^{\pi} = \frac{7}{2}^{-}\) for the 4142.9~keV level and \(J^{\pi} = \frac{9}{2}^{-}\) for the 4199.8~keV level.

The $\alpha$-decay branching ratios for these states have been reported in experiments that did not fully resolve the two levels. Laird~\textit{et al.}~\cite{2002La} established an upper limit of \(B_\alpha < 0.01\), whereas Tan~\textit{et al.}\cite{2009Ta} reported a combined value of \(B_\alpha = (1.2 \pm 0.5) \times 10^{-3}\). The average value is adopted for both levels in this evaluation. The total widths \(\Gamma\) were derived by converting the measured lifetimes from Refs.~\cite{1973Da, 2005Ta, 2008My}. The resulting evaluated widths are \(\Gamma = 4.1^{+0.8}_{-0.6} \times 10^{-2}~\mathrm{eV}\) for the 4142.8~keV level and \(\Gamma = (1.5 \pm 0.3) \times 10^{-2}~\mathrm{eV}\) for the 4199.8~keV level.

\textbf{4. E$_x$ = 4377.4$\pm$0.5 keV, $J^\pi$ = $\dfrac{7}{2}^+$}

The level at E$_x$ = 4377.4$\pm$0.5~keV is well established, with consistent spin-parity assignment of $J^\pi=\frac{7}{2}^+$ supported by multiple experimental studies~\cite{1971Bi,1973Da,1974Ga}. The mean lifetime \( \tau_m \) has been measured in three experiments~\cite{1973Da, 2005Ta, 2008My}. The evaluated total width is determined to be \( \Gamma = 0.16^{+0.07}_{-0.03}~\mathrm{eV} \).

Reported values of the \(\alpha\)-decay branching ratios \(B_\alpha\), however, exhibit significant variation as summarized in Table~\ref{tab:Ba measurements}. Magnus~\textit{et al.}~\cite{1990Ma} reported a relatively high value of \(B_\alpha = (4.4 \pm 3.2) \times 10^{-2}\), while Rehm~\textit{et al.}~\cite{2003Re} measured \( (1.6 \pm 0.5) \times 10^{-2} \). Davids~\textit{et al.}~\cite{2003Da} reported an upper limit of \( B_\alpha < 3.9 \times 10^{-3} \), while Visser~\textit{et al.}~\cite{2004Vi} presented a lower limit of \( B_\alpha > 2.7 \times 10^{-3} \). Tan \textit{et al.}~\cite{2009Ta} measured a much smaller value of \( (1.2 \pm 0.3) \times 10^{-3} \), and Bardayan \textit{et al.}~\cite{2019Ba} later reported an upper limit of \( B_\alpha < 3.0 \times 10^{-2} \).

The inconsistency among these results highlights the limitations of traditional averaging techniques. In this work, all reported values—including both upper and lower limits—are incorporated within a unified Bayesian framework without selective exclusion. The resulting evaluated value is \( B_\alpha = (2.1 \pm 0.2) \times 10^{-3} \), which reflects the full range of experimental constraints and associated uncertainties.

\textbf{5. E$_x$ = 4548.0$\pm$0.7~keV, $J^\pi = \dfrac{3}{2}^-$}

A spin-parity of \( J^{\pi} = \frac{3}{2}^{-} \) is assigned based on consistent experimental evidence from multiple studies~\cite{1970Ga,1971Bi,1973Da,2015Pa}. Although some earlier studies considered an alternative assignment of \( J^{\pi} = \frac{1}{2}^{-} \)~\cite{1970Ga,1971Bi,1973Da,2015Pa}, the mirror level in \(^{19}\mathrm{F} \) at \( E_x = 4556.1~\mathrm{keV} \)--firmly established as \( J^{\pi} = \frac{3}{2}^{-} \)--provides strong supporting evidence for the adopted assignment in \(^{19}\mathrm{Ne}\)~\cite{1995NDS}.

\textbf{6. E$_x$ = 4602.1$\pm$0.6~keV, $J^\pi = \dfrac{5}{2}^+$}

The level is assigned a spin-parity of \( J^{\pi} = \frac{5}{2}^{+} \), based on consistent experimental evidence from multiple studies~\cite{1973Da,2015Pa}. This assignment is further supported by the mirror level in \(^{19}\mathrm{F}\) at \( E_x = 4550~\mathrm{keV} \), which is firmly established as \( J^{\pi} = \frac{5}{2}^{+} \)~\cite{1995NDS}.

The \(\alpha\)-decay branching ratio for this level has been measured in several independent experiments. A combined analysis of results from Refs.~\cite{2004Vi,2009Ta,2003Da,2002La,1990Ma} yields an evaluated value of \( B_\alpha = 0.26 \pm 0.01 \). This result is in excellent agreement with the weighted average of the individual measurements, demonstrating good consistency and illustrating the compatibility between the Bayesian and conventional evaluation approaches (See Sec.~\ref{sec:advantage}).

\textbf{7. E$_x$ = 4633.8$\pm$0.6~keV, $J^\pi = \dfrac{13}{2}^+$}

The state is assigned \( J^{\pi} = \frac{13}{2}^{+} \), consistent with the mirror level in \(^{19}\mathrm{F}\) at \(4648~\mathrm{keV} \)~\cite{1995NDS,1971Bi,1972Pa,1973Da}. Two independent measurements of the mean lifetime $\tau_m$ both reported values greater than 1000~fs, from which an upper limit on the total width of $\Gamma<1.05\times10^{-3}$~eV is inferred~\cite{1973Da,2005Ta}. The $\alpha$-branching ratio B$_\alpha$ adopted in this work is taken from Ref.~\cite{1995NDS}.

\textbf{8. E$_x$ = 4708.4$\pm$1.7 keV, $J^\pi = \dfrac{5}{2}^-$}

The spin-parity of the state is assigned as $\frac{5}{2}^-$~\cite{2015Pa}.
The branching ratio $B_\alpha=0.82\pm0.03$ has been consistently reported~\cite{1990Ma,2003Da,2009Ta,2004Vi}, while other resonance parameters remain unmeasured. The mean lifetime $\tau_m$ is indirectly constrained using the mirror level in $^{19}$F~\cite{1995NDS}.

\textbf{9. E$_x$ = 5090.8$\pm$1.9 keV, $J^\pi = \dfrac{5}{2}^+$}

The ninth state above the $\alpha$-threshold is located at E$_x$ = 5090.8$\pm$1.9~keV and is consistently assigned as a spin-parity of $J^\pi=\frac{5}{2}^+$~\cite{1971Bi,2015Ba}. The evaluated $\alpha$-decay branching ratio is $B_\alpha=0.82\pm0.02$. Since the value of $\tau_m$ has not been measured directly, the corresponding value from the mirror state in $^{19}$F is adopted.

\textbf{10-12. E$_x$ = 5351$\pm$3 keV \& 5424$\pm$5 keV \& 5488$\pm$3 keV, $J^\pi = \dfrac{1}{2}^+$ \& $\dfrac{7}{2}^+$ \& $\dfrac{3}{2}^+$}

The tenth state above the $\alpha$-threshold has an excitation energy of 5351$\pm$3~keV. The spin-parity is assigned as $\frac{1}{2}^+$ based on multiple consistent observations~\cite{1971Bi,1970Ga,2017Ka,2017To,2022Go}. The $\alpha$-decay width $\Gamma_\alpha$ has been measured in two studies~\cite{2006Va,2017To} and is evaluated as 4.4$\pm$1.5~keV.

The eleventh state is located at E$_x$ = 5424$\pm$5~keV and is assigned $J^\pi=\frac{7}{2}^+$~\cite{1971Bi,1972Pa}. Since neither its decay width nor lifetime has been measured experimentally, the total width is estimated from the mirror state in $^{19}$F at E$_x$ = 5463.5~keV, which has a reported lifetime of 0.26~fs~\cite{1995NDS}. This corresponds to an evaluated total width of $\Gamma<8.9$~eV.

The twelfth state lies at E$_x$ = 5488$\pm$3~keV with an assigned spin-parity of $J^\pi=\frac{3}{2}^+$~\cite{2017Ka,2017To,2022Go}. Its $\alpha$-decay width has been measured in Ref.~\cite{2017To} and is evaluated as $\Gamma_\alpha=9\pm$2~keV.

The $\alpha$-decay branching ratios $B_\alpha$ for these three levels have been measured only once~\cite{2002La}. In that experiment, the levels at E$_x$ = 5351, 5424, and 5488~keV could not be resolved, and a combined branching ratio of $B_\alpha=1.3\pm$0.3 was reported. In the present evaluation, a lower limit of $B_\alpha>0.66$ is extracted by combining a Gaussian distribution based on the measured value with a uniform prior. Additional experimental data are needed to better constrain the individual branching ratios for these states.

\textbf{13. E$_x$ = 5535$\pm$7 keV, $J^\pi = \dfrac{5}{2}^+$ or $\dfrac{7}{2}^-$}

The level has not been definitively assigned a spin-parity from experimental data. Therefore, the possible $J^\pi$ values ($\frac{5}{2}+$ or $\frac{7}{2}-$) are inferred from the corresponding mirror level in $^{19}$F. The recommended $p(J^\pi)$ for each assignment is 0.5. Notably, more excited states have been identified in $^{19}$F than in $^{19}$Ne, suggesting that this level in $^{19}$Ne could represent a superposition or unresolved multiplet of several states. The $\alpha$-branching ratio has been tentatively measured as $B_\alpha$=0.88$\pm$0.05 in Ref.~\cite{2019Ba}. The $\alpha$-decay width $\Gamma_\alpha$ in Table~\ref{tab:Tab3} is derived using the resonance parameters of the potent mirror state at E$_x$ = 5418~keV in $^{19}$F ($J^{\pi}$=$\frac{7}{2}^-$).

\textbf{14. E$_x$ = 5704$\pm$8 keV, $J^\pi = \dfrac{5}{2}^-$}

The fourteenth state at E$_x$ = 5704$\pm$8~keV is assigned $J^\pi=\frac{5}{2}^-$ based on Ref.~\cite{2017To}. This level has been reported in only one experiment~\cite{2017To}, in which the $\alpha$-decay width was measured as $\Gamma_\alpha=29\pm$6~keV. The evaluated $\alpha$-decay width is 27$\pm$6~keV. The total width $\Gamma$ is calculated using the mean lifetime of the corresponding mirror state in $^{19}$F at E$_x$ = 5621~keV~\cite{2017To,1995NDS}, for which an upper limit of $\tau_m\leq{1.3}$~fs has been reported.

\textbf{15. E$_x$ = 5830$\pm$6 keV, $J^\pi = \dfrac{1}{2}^+$ or $ \dfrac{3}{2}^+$}

The fifteenth spin-parity of the state is constrained to $J^\pi=\frac{1}{2}^+$ or $\frac{3}{2}^+$~\cite{2017Ka,1995NDS}. The recommended $p(J^\pi)$ is estimated to be 0.75 and 0.5, respectively. Experimental measurements of its resonance parameters have scarcely been reported since its first observation in Ref.~\cite{1970Ga}. The resonance parameters of the corresponding mirror state at E$_x$ = 5938~keV in $^{19}$F remain largely unknown, except for its $\alpha$-decay width. An R-matrix analysis in Ref.~\cite{2019Co} estimated $\Gamma_\alpha=0.77~$keV for the mirror state. The probability distribution for $\Gamma_\alpha$ of the E$_x$ = 5830~keV state in $^{19}$Ne is derived following the procedure described in Sec.~\ref{sec: tau_gamma}.

\subsubsection{Above E$_x$ = 6014~keV, below E$_x$ = 6417~keV}
This energy region is important for the $^{18}$F($p, \alpha$)$^{15}$O reaction rate, owing to the presence of a subthreshold resonance that may strongly influence the rate at low temperatures (T$_9\leq0.3$~GK). Due to the greater uncertainty in mirror-state assignments at these excitation energies, only experimentally determined parameters are used in the present evaluation. Although not evaluated in this work, Asymptotic Normalization Coefficient (ANC) values reported in previous experimental studies are presented for levels where available, as they are important for calculating the contributions of subthreshold resonances.

\begin{table*}
\caption{Evaluated decay information for levels in $^{19}$Ne including potential subthreshold resonance levels that contribute to the $^{18}$F($p, \alpha$)$^{15}$O reaction. The evaluated spin-parity assignments are presented. The references for spin-parity are in Table~\ref{tab: pJ}.}\label{tab:Tab3}
\begin{ruledtabular}

\begin{tabular}{ccccccc}
\multirow{2}{*}{E$_x$ [keV]} &\multirow{2}{*}{$J^\pi$} &\multirow{2}{*}{$B_\alpha$} &\multirow{2}{*}{$\Gamma_\alpha$ [keV]} &\multicolumn{3}{c}{Reference} \\\cmidrule{5-7}
& & & &E$_x$ &$B_\alpha$ &$\Gamma_\alpha$ \\
\hline
6014$\pm$2& $\frac{3}{2}^-$&$>0.69$& $17^{+8}_{-9}$&\cite{1970Ga,1972Ha,2013La,2015Ba,2017Ka,2017To,2022Go}&\cite{2002La,2019Ba}&~\cite{2017To}\\
6081$\pm$1& $\frac{5}{2}^-/\frac{3}{2}^+$&$>0.69$& &\cite{1970Ga,1979Ma,2011Ad(b),2013La,2015Ch}&\cite{2002La,2019Ba}& \\
6100$\pm$1& $\frac{7}{2}^+$&$>0.69$& &\cite{1972Ha,2013La,2015Ba,2020Ha}&\cite{2002La,2019Ba}& \\
6138$\pm$2& $\frac{1}{2}^+/\frac{3}{2}^+$&& $15\pm5$&\cite{2013La,2017Ka,2017To,2022Go,1970Ga}&&\cite{2017To}\\
(6272)& $\frac{7}{2}^+$&& (5.5)&\cite{2022Go}&& \cite{2022Go}\\
(6279$\pm$2)&$\frac{5}{2}^+$&& ($5\pm2$)&\cite{2017To}&& \cite{2017To}\\
6286$\pm$1&$\frac{1}{2}^+/\frac{3}{2}^+$&$>0.82$& &\cite{2011Ad(b),2013La,2015Ba,2015Ch,2015Pa,2017Ka,1979Ma,1970Ga,1972Ha,1972Pa}&\cite{2019Ba,2021Ri}& \\
6292$\pm$1&$\frac{11}{2}^+$&& &\cite{2020Ha,2015Pa}&& \\
\end{tabular}
\end{ruledtabular}
\end{table*}

\textbf{16-18. E$_x$ = 6014$\pm$2 keV \& 6081$\pm$1 keV \& 6100$\pm$1 keV, $J^\pi = \dfrac{3}{2}^-$ \& $\dfrac{5}{2}^-$ or $\dfrac{3}{2}^+$ \& $\dfrac{7}{2}^+$}

These three levels form a triplet near $E_x \approx 6.0$~MeV as reported in Ref.~\cite{2013La}. The E$_x$ = 6014~keV state is assigned $J^\pi = \frac{3}{2}^-$ based on multiple experimental observations~\cite{2022Go,2017To,2013La,1972Ha,1970Ga}. For the E$_x$ = 6081~keV state, the spin-parity remains uncertain; angular momentum transfer analysis suggests either $J^\pi = \frac{5}{2}^-$ or $\frac{3}{2}^+$~\cite{2013La}. Accordingly, an equal probability of $p(J^\pi) = 0.5$ is recommended for each assignment. The E$_x$ = 6100~keV level is assigned $J^\pi = \frac{7}{2}^+$~\cite{2013La,2020Ha}. The $\alpha$-branching ratios for the triplet have been measured in Refs.~\cite{2002La,2019Ba}, although the individual levels were not resolved in either study. Laird~\textit{et al.}~\cite{2002La} reported $B_\alpha = 0.96 \pm 0.20$ for a doublet assumed to correspond to the E$_x = 6013$ and 6092~keV states, with the latter likely representing an unresolved blend of the 6081 and 6100~keV levels. Combining the available data, the $\alpha$-branching ratio for the full triplet is evaluated as $B_\alpha > 0.69$ at 90\% credible level.

Riley~\textit{et al.}~\cite{2021Ri} tentatively reported a broad state at E$_x$ = 6008~keV with a total width of $\Gamma$ = 124$\pm$25~keV, suggesting that it may influence on the $^{18}$F$(p, \alpha)^{15}$O reaction rate through a broad subthreshold resonance. However, in that study, the nearby triplet levels were not fully resolved, introducing significant uncertainty into the extracted width. Consequently, the reported $\Gamma$ value is not considered in the present evaluation.

\textbf{19. E$_x$ = 6138$\pm$2 keV, $J^\pi = \dfrac{1}{2}^+$ or $\dfrac{3}{2}^+$}

The state is a candidate for a broad subthreshold resonance state with ${\Delta}l = 0$~\cite{2017Ka}, potentially influencing the $^{18}$F($p, \alpha$)$^{15}$O reaction rate. The spin-parity remains uncertain, with assignments of $J^\pi=\frac{1}{2}^+$ or $\frac{3}{2}^+$ reported in Refs.~\cite{2013La,2017To,2017Ka,2022Go,2023Po}. Although earlier study~\cite{2017Ka} suggested ${\Delta}l = 0$, a more recent DWBA analysis~\cite{2023Po} indicates ${\Delta}l = 2$, favoring a higher angular momentum transfer. Consequently, the recommended $p(J^\pi)$ is 0.5 for each assignment. The $\alpha$-decay width has been measured once in Ref.~\cite{2017To} and is evaluated as $\Gamma_\alpha=15\pm5$~keV. Kahl~\textit{et al.} \cite{2017Ka} estimated the ANC for this level as 8~fm$^{-1/2}$ for $\frac{1}{2}^+$, or 6~fm$^{-1/2}$ for $\frac{1}{2}^+$; however, associated uncertainties were not reported. To better constrain the contribution of this level to the reaction rate, future measurements for the ANC are needed.


\textbf{(E$_x$ = 6272 keV \& 6279$\pm$2~keV, $J^\pi = \dfrac{7}{2}^+$ \& $\dfrac{5}{2}^+$)}

The states at E$_x$ = 6271~keV and 6279~keV were reported in Refs.~\cite{2017To,2022Go}. As the work in Ref.~\cite{2022Go} is a reanalysis of the experimental data originally presented in Ref.~\cite{2017To}, both states effectively originate from a single experimental investigation. Independent confirmation is therefore required to establish the existence of these levels with confidence. 

\textbf{20-21. E$_x$ = 6286$\pm$1 \& 6292$\pm$1~keV, $J^\pi = \dfrac{1}{2}^+$ or $\dfrac{3}{2}^+$ \& $\dfrac{11}{2}^+$}

The states at E$_x$ = 6286$\pm$1 and 6292$\pm$1~keV likely form a doublet~\cite{2023Po,2021Ri}. The E$_x$ = 6286~keV level is considered a strong candidate for a subthreshold s-wave resonance, with tentatively assigned spin-parity of $J^\pi=\frac{1}{2}^+$ or $\frac{3}{2}^+$~\cite{2011Ad(b),2013La,2015Ba}. The recommended $p(J^\pi)$ is equally 0.5 for both assignments. The ANC for this state has been determined as C=3479$\pm$92 fm$^{-1}$ for $\frac{3}{2}^+$, or 6972$\pm$183~fm$^{-1}$ for $\frac{1}{2}^+$~\cite{2011Ad}. The 6292~keV state has been proposed as a high-spin $J^\pi=\frac{11}{2}^+$ level, originally predicted through mirror analysis~\cite{2007Ne} and later supported by a $\gamma$-ray measurement~\cite{2020Ha}. The $\alpha$-decay branching ratio for this doublet has been measured twice in Ref.~\cite{2021Ri} and \cite{2019Ba}.

Based on mirror symmetry, Nesaraja \textit{et al.}~\cite{2007Ne} predicted the existence of a high-spin state in $^{19}$Ne with $J^\pi = \frac{11}{2}^+$ near E$_x$ = 6422$\pm$30~keV. This prediction was supported by Laird~\textit{et al.}~\cite{2013La}, who observed a state at E$_x$ = 6440~keV and assigned $J^\pi = \frac{11}{2}^+$. However, this assignment has not been confirmed by other experiments, and the state at E$_x$ = 6440~keV has also been proposed as the mirror of the $^{19}$F level at E$_x$ = 6536~keV with $J^\pi=\frac{1}{2}^-$~\cite{1998Uk,2007Ne,2000Co} (See section below).

Several studies have reported unresolved features in this energy region. Laird~\textit{et al.}~\cite{2013La} observed angular distribution curves indicative of an unresolved doublet at E$_x$ = 6289~keV, while Parikh~\textit{et al.}~\cite{2015Pa} proposed a doublet at E$_x$ = 6282 and 6295~keV. Further evidence for a high-spin component was provided by Hall \textit{et al.} \cite{2020Ha}, who detected $\gamma$-ray transitions from a state at E$_x$ = 6291.6~keV, consistent with a high-spin assignment. Although $\gamma$-ray transitions from the low-spin member of the doublet were not observed in that study, more recent measurements~\cite{2021Ri,2023Po} have reported angular distributions consistent with a mixed population from low- and high-spin states. While the doublet has not yet been fully resolved experimentally, the available evidence strongly supports its existence.

\subsubsection{From E$_x$ = 6417~keV (S$_p$)}

The energy levels in this region lie within the Gamow window for the $^{18}\mathrm{F}(p,\alpha)^{15}\mathrm{O}$ reaction at stellar temperatures $T_9 \approx 0.4$~GK. These levels are particularly important for reaction rate calculations due to the potential influence of broad subthreshold resonances, interference effects, and narrow resonances near the proton threshold. Numerous experimental and theoretical studies have highlighted the sensitivity of the reaction rate to the properties of these states~\cite{2006Chae,2011Ad,2013La,2021Ka}. In this section, we present the evaluated resonance parameters for levels from E$_x = 6417$~keV, with particular attention given to those known to significantly impact the astrophysical reaction rate. States with less pronounced influence are summarized briefly.

\begin{turnpage}
\begin{table*}
\caption{Evaluated resonance parameters in levels that contribute to the $^{18}$F($p, \alpha$)$^{15}$O reaction in stellar temperatures ($\approx0.4$~GK) and the references are tabulated. For more details, see text. The recommended spin-parity assignments are presented.}\label{tab:Tab4}
\begin{ruledtabular}

\renewcommand{\arraystretch}{1.5}
\scriptsize
\begin{tabular}{ccccccccccc}
\multirow{2}{*}{E$_x$ [keV]} &\multirow{2}{*}{$J^\pi$} &\multirow{2}{*}{$B_\alpha$} &\multirow{2}{*}{$\Gamma$ [eV]} &\multirow{2}{*}{$\Gamma_\alpha$ [keV]} &\multirow{2}{*}{$\Gamma_p$ [keV]} &\multicolumn{5}{c}{Reference}\\
\cmidrule{7-11}
& & & & & &E$_x$ &$B_\alpha$ &$\Gamma$ &$\Gamma_\alpha$&$\Gamma_p$ \\
\hhline{-----------}
6417$\pm$2& $\frac{3}{2}^-$& & & & $(1.27\pm0.04)\times10^{-38}$&\cite{2011Ad(b),2013La,2017Ka,1998Uk}& & & & \cite{2011Ad}\\
6423$\pm$3& $\frac{3}{2}^+$& & & & &\cite{2020Ha}& & & & \\
6435$\pm$2& $\frac{1}{2}^-$& $>0.89$& $216\pm19$& $108^{+62}_{-63}$& &\cite{1972Ha,1970Ga,1998Uk,2017To,2013La,2015Ba,2017To}& \cite{2019Ba}& \cite{1998Ut}& \cite{2022Go}&\\
6450$\pm$2& $\frac{3}{2}^+$& & & & $<1.89\times10^{-15}$&\cite{2013La,2015Ch,2020Ha,1998Uk}& & & & \cite{2011Ad(b)}\\
(6537$\pm$15)& $\frac{7}{2}^+$& & & & &\cite{2015Ch,2022Go}& & & & \\
6700$\pm$3& $\frac{5}{2}^-/\frac{5}{2}^+$& & & $24^{+2}_{-3}$& &\cite{2013La,1998Ut,2022Go}& & & \cite{2022Go}&\\
6743$\pm$2& $\frac{3}{2}^-$& $0.95^{+0.3}_{-0.4}$& & $4.2\pm0.4$& $5.1^{+0.04}_{-0.05}\times10^{-3}$&\cite{2011Ad(b),2013La,2015Ba,2015Ch,2017Ka,2020Ha,1998Ut,1970Ga,1972Ha,2022Go}& \cite{1998Ut,2004Vi,2021Ri}& & \cite{2022Go}& \cite{2011Ad,2002Ba}\\
(6851$\pm$4)& $\frac{5}{2}^+$& & & & &\cite{2015Pa,1972Pa}& & & &\\
6863$\pm$1& $\frac{7}{2}^-$& $0.92\pm0.3$& & & &\cite{2013La,2015Ba,2015Pa,2020Ha,1979Ma,1998Ut,1970Ga,1972Ha,2022Go}& \cite{1998Ut,2004Vi,2021Ri}& & & \\
6967$\pm$19& $\frac{5}{2}^+$& & & $28^{+2}_{-3}$& &\cite{2015Ch,2022Go}& & & \cite{2022Go}&\\
(7027)& $\frac{1}{2}^-$& & & $141\pm14$& &\cite{2022Go}& & & \cite{2022Go}& \\
7072$\pm$1& $\frac{3}{2}^+$& $0.613\pm0.012$& $39.07\pm1.50$& ($15\pm1$)& $13.8\pm2.0$&\cite{2009Da,2009Mu,2011Ad(b),2015Ba,2015Ch,2017Ka,2022Go,1998Ut,1970Ga,1972Ha}& \cite{1998Ut,2004Vi,2021Ri,2001Ba}& \cite{2017Ka,1998Ut,2009Da,2001Ba}& \cite{2009Mu,2022Go}&\cite{2009Mu,2011Ad}\\
7173$\pm$2& $\frac{11}{2}^-$& & & & &\cite{2015Pa,1998Ut,1972Ha}& & & &\\
7230$\pm$5& $\frac{3}{2}^+$& & & $<7.9$& $<0.6$&\cite{2011Ad(b),1979Ma,1998Ut,1972Pa,2012Mo}& & & \cite{2012Mo}& \cite{2012Mo,2004Ba_19Ne}\\
7279$\pm$7& $\frac{1}{2}^+$& & $35\pm12$& $35\pm3$& $<1.5$&\cite{2012Ad,2015Ch,2022Go,1972Ha,2009Da}& & \cite{2009Da}& \cite{2022Go}&\cite{2004Ba_19Ne}\\
7396$\pm$5& $\frac{7}{2}^+$& $0.76\pm0.12$& & $99\pm7$& $27\pm4$&\cite{2011Ad(b),2017To,1972Ha}& \cite{2021Ri}& & \cite{2017To,2004Ba_19Ne}& \cite{2004Ba_19Ne}\\
7495$\pm$3& $\frac{5}{2}^+$& $0.19\pm0.02$& $17\pm6$& $1.8\pm1.1$& $1.7^{+0.5}_{-0.7}$&\cite{2009Da,2009Mu,2015Ch,2017To,1998Ut,2012Mo}& \cite{2021Ri,1998Ut}& \cite{1998Ut,2009Da}& \cite{2009Mu,2012Mo,2017To}& \cite{2009Mu,2012Mo,2004Ba_19Ne}\\
7532$\pm$9& $\frac{5}{2}^-$& $0.67\pm0.08$& $31\pm16$& & &\cite{1998Ut,1972Ha,1972Pa}& \cite{1998Ut}& \cite{1998Ut}& & \\
7615$\pm$3& $\frac{3}{2}^+$& $0.96^{+0.03}_{-0.04}$& $28^{+8}_{-9}$& $23.6^{+0.4}_{-0.3}$& $2.5^{+0.6}_{-0.5}$&\cite{2009Da,2017Ka,2017To,1998Ut,1972Ha,2012Mo,2012Ad}& \cite{1998Ut}& \cite{1998Ut,2009Da}& \cite{2009Mu,2012Mo,2017To}& \cite{2009Mu,2012Mo}\\
7668$\pm$7& $\frac{3}{2}^-$& $0.37\pm0.06$& $43\pm16$& $<0.2$& $2.7\pm1.0$&\cite{2009Mu,2022Go,1998Ut,1972Ha,2012Mo}& \cite{1998Ut}& \cite{1998Ut}& \cite{2022Go,2009Mu,2012Mo}& \cite{2009Mu,2012Mo}\\
7764$\pm$3& $\frac{3}{2}^+$&$0.19\pm0.09$& $100^{+8}_{-9}$& $7.3\pm0.7$& $55^{+7}_{-6}$&\cite{2009Mu,2017Ka,2022Go,1998Ut,1972Ha,2012Mo}& \cite{1998Ut}& \cite{2017Ka,1998Ut}& \cite{2022Go,2009Mu,2012Mo}& \cite{2009Mu,2012Mo}\\
7872$\pm$19& $\frac{1}{2}^+$& & $292\pm107$& $237\pm23$& $62\pm12$&\cite{2009Da,2011Ad(b),2022Go,2012Mo}& & \cite{2009Da}& \cite{2022Go,2012Mo}&\cite{2012Ad,2012Mo}\\
7995$\pm$5& $\frac{5}{2}^-$& & $12\pm7$& $12\pm2$& $1.9^{+0.3}_{-0.4}$&\cite{2009Da,2009Mu,2022Go,1968Gu,1971Bi,1972Ha,1972Pa,2012Mo}& & \cite{2009Da}& \cite{2022Go,2009Mu,2012Mo}& \cite{2009Mu,2012Mo}\\
\end{tabular}
\end{ruledtabular}

\end{table*}
\end{turnpage}

\textbf{22-23. E$_x$ = 6417$\pm$2 keV \& 6423$\pm$3 keV, $J^\pi = \dfrac{3}{2}^-$ \& $\dfrac{3}{2}^+$}

The twenty-second and-third excited states have been assigned spin-parities $J^\pi = \frac{3}{2}^-$ and $\frac{3}{2}^+$. Positioned just above the proton threshold, the states have been extensively studied due to their significant impact on the $^{18}\mathrm{F}(p,\alpha)^{15}\mathrm{O}$ reaction rate. The 6417~keV state has been assigned $J^\pi = \frac{3}{2}^-$ based on angular distribution analyses from transfer reactions such as $^{18}\mathrm{F}(d,n)^{19}\mathrm{Ne}$ or $^{19}$F($^3He, t)^{19}$Ne~\cite{2011Ad,2013La}. Laird~\textit{et al.}~\cite{2013La} observed this state and tentatively proposed assignment of $J^\pi =\frac{3}{2}^-$ or $\frac{5}{2}^+$. The 6423~keV state has been assigned $J^\pi = \frac{3}{2}^+$, supported consistently by $\gamma$-ray spectroscopy and transfer reaction data~\cite{2017Ka,2020Ha}. Although these two states have not yet been fully resolved experimentally, recent evaluations generally treat them as distinct levels~\cite{2019Ha_PRL,2021Ka}.

These near-threshold states are important due to the potential interference effects among resonances sharing the same spin-parity, particularly the $\frac{3}{2}^+$ states~\cite{2021Ka,2006Chae}. In addition, narrow-isolated resonances involving these low-spin states can significantly influence reaction rates. Hence, the charged-particle partial widths --- particularly $\Gamma_p$ --- of these levels are critical parameters for accurate reaction rate calculations. Although indirect estimates of $\Gamma_p$ exist~\cite{2002Ba, 2013La}, further experimental constraints are required. For instance, Bardayan~\textit{et al.}) \cite{2002Ba} calculated $\Gamma_p=(3.9\pm3.9)\times10^{-37}$~keV for E$_x$ = 6423~keV ($J^\pi=\frac{3}{2}^+$) with the reduced proton width adopted from the analog level~\cite{2003Se,2005Ko}. Laird \textit{et al}.~\cite{2013La} recalculated $\Gamma_p=4.7\times10^{-50}$~keV for E$_x$ = 6417~keV ($J^\pi=\frac{3}{2}^-$) state.

\textbf{24. E$_x$ = 6435$\pm$2 keV, $J^\pi = \dfrac{1}{2}^-$ or $\dfrac{11}{2}^+$}

The twenty-fourth level has been discussed with two possible spin-parity assignments: $J^\pi=\frac{1}{2}^-$ and $\frac{11}{2}^+$. The assignment of $J^\pi=\frac{1}{2}^-$ is supported by a $\alpha-$scattering experiment \cite{2017To,2022Go}, and is consistent with the mirror state at E$_x = 6536$~keV in $^{19}$F~\cite{2007Ne}. However, one angular distribution measurement from the $^{19}\mathrm{F}(^{3}\mathrm{He},t)^{19}\mathrm{Ne}$ reaction suggests a high-spin assignment of $J^\pi=\frac{11}{2}^+$, proposing this as a candidate mirror of the $^{19}$F level at $E_x = 6500$~keV~\cite{2013La}. Despite this ambiguity, the $J^\pi = \frac{1}{2}^-$ assignment is currently favored due to broader consistency with the systematics of mirror nuclei. Consequently, the recommended $p(J^\pi)$ is 0.75 for $J^\pi=\frac{1}{2}^-$ and 0.25 for $\frac{11}{2}^+$.

The total width ($\Gamma$), the $\alpha$-decay width ($\Gamma_\alpha$), and the $\alpha$-branching ratio ($B_\alpha$) have been reported in Refs.~\cite{1998Ut,2017To,2022Go,2019Ba}. The proton partial width ($\Gamma_p$), however, has not been measured directly. As an interim measure, theoretical estimates based on reduced widths and mirror-level structure assumptions have been adopted in studies such as Ref.~\cite{2015Ba}.

\textbf{25. E$_x$ = 6450$\pm$2 keV, $J^\pi = \dfrac{3}{2}^+$ or $\dfrac{5}{2}^-$}



The twenty-fifth level has two proposed spin-parity assignments: $J^\pi=\frac{3}{2}^+$ or $\frac{5}{2}^-$. The $J^\pi = \frac{3}{2}^+$ assignment is currently favored, supported by multiple experimental studies involving $\gamma$-ray spectroscopy and Trojan horse method~\cite{2020Ha,2015Ch}. Although Laird~\textit{et al.} \cite{2013La} tentatively suggested $J^\pi = \frac{5}{2}^-$ assignment, subsequent experiments have provided limited support for this interpretation. For example, no mirror state with $J^\pi=\frac{5}{2}^-$ has been identified near E$_x$ = 6450~keV~\cite{2019Co,1995NDS,2005Ba}. Hence, the recommended $p(J^\pi)$ is 0.75 for $J^\pi=\frac{3}{2}^+$, and 0.25 for $\frac{5}{2}^-$.

This state lies within the Gamow window and therefore plays a significant role in interference effects among nearby $J^\pi=\frac{3}{2}^+$ resonances, strongly influencing the $^{18}\mathrm{F}(p,\alpha)^{15}\mathrm{O}$ reaction rate at elevated stellar temperatures. However, direct measurements of key resonance parameters such as $\Gamma_\alpha$ are currently unavailable. Future experimental work is essential to clarify the spin-parity assignment and to accurately determine the partial widths necessary for reliable reaction rate calculations.

\textbf{(E$_x$ = 6537$\pm$15 keV, $J^\pi = \dfrac{7}{2}^+$)}

This state has been considered a missing level, initially assumed to lie at E$_x$ = 6504$\pm$30~keV as the mirror counterpart of the $^{19}$F state at E$_x$ = 6554~keV with $J^\pi=\frac{7}{2}^+$~\cite{2007Ne}. Cherubini~\textit{et al.} \cite{2015Ch} observed a peak corresponding to a level at E$_x$ = 6537~keV, assigning it a tentative spin-parity of either $J^\pi=\frac{7}{2}^+$ or $\frac{9}{2}^+$. However, this state has not been consistently observed by other experimental approaches, such as transfer reaction or $\gamma$-ray spectroscopy, and its existence remains unconfirmed.

\textbf{26. E$_x$ = 6700$\pm$3 keV, $J^\pi = \dfrac{5}{2}^+$ or $\dfrac{5}{2}^-$}

The twenty-sixth level has a tentative spin-parity assignment of $J^\pi=\frac{5}{2}^+$ or $\frac{5}{2}^-$. The $\alpha$-decay width ($\Gamma_\alpha$) of this level was determined from an $\alpha$-scattering experiment~\cite{2022Go}. The $J^{\pi}$ assignment is inferred from the presumed mirror state~\cite{2005Ko} and further supported by an R-matrix analysis presented in Ref.~\cite{2022Go}. The recommended $p(J^\pi)$ is equally 0.5 for each assignment.

Other resonance parameters have not been directly measured. However, Bardayan~\textit{et al.} \cite{2002Ba} recalculated the proton decay width ($\Gamma_p$) by adopting parameters from the corresponding mirror state reported by Kozub~\textit{et al.} \cite{2005Ko}.

\textbf{27. E$_x$ = 6743$\pm$2 keV, $J^\pi = \dfrac{3}{2}^-$}

The twenty-seventh level has an excitation energy of E$_x$ = 6743$\pm$2~keV with $J^{\pi}=\frac{3}{2}^-$. This level has been identified as a key resonance that significantly influences the $^{18}$F($p, \alpha$)$^{15}$O reaction rates~\cite{2002Ba}.

The assignment of $J^\pi=\frac{3}{2}^-$ has been consistently supported by multiple experimental studies, as tabulated in Table~\ref{tab: pJ}. The proton decay width ($\Gamma_p$) has been measured in two independent experiments, yielding values of (2.22$\pm$0.69)$\times$10$^{-3}$~keV in Ref.~\cite{2002Ba} and (7.3$\pm0.6)\times10^{-3}$~keV in Ref.~\cite{2011Ad}. The present evaluation provides an updated value of $\Gamma_p=5.1^{+0.04}_{-0.05}\times10^{-3}$~keV. The $\alpha$-decay width ($\Gamma_\alpha$), deduced from an R-matrix analysis in Ref.~\cite{2022Go}, is evaluated and presented in Table~\ref{tab:Tab4}.

\textbf{28. (E$_x$ = 6851$\pm$4 keV, $J^\pi = \dfrac{5}{2}^+$) \& E$_x$ = 6863$\pm$1, $J^\pi = \dfrac{7}{2}^-$}

The twenty-eighth level has an excitation energy of E$_x$ = 6863$\pm$1~keV with $J^{\pi}=\frac{7}{2}^-$. The $\alpha$-decay branching ratio (B$_\alpha$) for this state has been measured in three independent experiments~\cite{1998Ut,2021Ri,2004Vi}, while other resonance parameters remain undetermined. The spin-parity assignment of the E$_x$ = 6863~keV state is adopted from Refs.~\cite{2013La,2023Po} and further supported by comparison with its mirror state in $^{19}$F, as listed in the Ref.~\cite{1995NDS}.

Additionally, the triton spectra from the $^{19}$F($^3He, t$)$^{19}$Ne reaction reported in Ref.~\cite{2015Pa} suggest the presence of a doublet structure consisting of E$_x$ = 6851 and 6863~keV states. Considering experimental studies and mirror symmetry~\cite{1972Pa,2015Pa,1995NDS}, a probability weight of 0.5 is assigned to the $J^\pi = \frac{5}{2}^+$ assignment for the E$_x$ = 6851~keV state. The alternative spin-parity values proposed by Panagiotou and Gove~\cite{1972Pa} are assigned a weight of 0.125 each.

\textbf{(E$_x$ = 6967$\pm$19 keV \& E$_x$ = 7027 keV, $J^\pi = \dfrac{5}{2}^+$ \& $\dfrac{1}{2}^-$)}

These two levels were initially predicted through mirror symmetry analysis in Ref.~\cite{2007Ne} and later reported in Refs.~\cite{2015Ch,2022Go}. The three-body reaction $^2H(^{18}F, \alpha^{15}O)n$ was studied utilizing the Trojan horse method in Ref.~\cite{2015Ch}. In the resulting cross section spectrum (see Fig.~3 of Ref.~\cite{2015Ch}), a peak corresponding to the E$_x$ = 6967~keV state was identified. The spin-parity of this level was later assigned based on $\alpha$-scattering measurements described in Ref.~\cite{2022Go}, which also reported an additional state at E$_x$ = 7027~keV with $J^\pi=\frac{1}{2}^-$. However, the existence of both states has not been consistently confirmed by other experimental investigations.

\textbf{29. E$_x$ = 7072$\pm$1 keV, $J^\pi = \dfrac{3}{2}^+$}

The state is located at E$_x$ = 7072$\pm$1~keV with $J^\pi=\frac{3}{2}^+$. This level is considered one of the most likely candidates to interfere with nearby $J^\pi=\frac{3}{2}^+$ resonances~\cite{2006Chae,2009Se,2023Po,2019Ha_PRL}, potentially impacting the $^{18}$F($p, \alpha$)$^{15}$O reaction rate. Its resonance parameters have been extensively investigated and are summarized in Table~\ref{tab:Tab4}.

\textbf{30. E$_x$ = 7173$\pm$2 keV, $J^\pi = \dfrac{11}{2}^-$}

The level has a proposed spin-parity of $J^\pi=\frac{11}{2}^-$. The assignment is based on mirror symmetry considerations~\cite{2007Ne} and is supported by the DWBA analysis presented in Ref.~\cite{2015Pa}. In contrast, Mountford~\textit{et al.} \cite{2012Mo} reported a resonance at E$_{c.m.}$=0.759~MeV with $J^\pi=\frac{3}{2}^+$, which contradicts the interpretation of Ref.~\cite{2015Pa}. However, the triton energy spectrum shown in Ref.~\cite{2015Pa} clearly resolved a peak corresponding to the E$_x$ = 7238~keV state, indicating that the 7173~keV state was not responsible for the observed feature in the Mountford study. Therefore, the spin-parity assignment from Ref.~\cite{2015Pa} is adopted in the present evaluation.

\textbf{31-32. E$_x$ = 7230$\pm$5 keV \& 7279$\pm$7 keV, $J^\pi = \dfrac{3}{2}^+$ \& $\dfrac{1}{2}^+$ or $\dfrac{3}{2}^+$}

The thirty-first state at E$_x$ = 7230$\pm$5~keV has a proposed spin-parity of $J^\pi=\frac{3}{2}^+$. The level was initially considered a potential interference partner of the E$_x$ = 7072~keV ($\frac{3}{2}^+$) state. However, Sereville~\textit{et al.} \cite{2009Se} demonstrated that the interference between these two states has a negligible effect on the reaction rate in the energy range (E$_{c.m}<400~\text{keV}$) relevant to nova conditions. Accordingly, this state has not been emphasized in recent evaluations of the reaction rate.

The thirty-second state at E$_x$ = 7279$\pm$7~keV with proposed spin-parity of $J^\pi=\frac{1}{2}^+$ remains less well established. The $\frac{1}{2}^+$ and $\frac{3}{2}+$ assignments are given weights of 0.75 and 0.25, respectively. Its existence has not been definitely confirmed, as it has not been observed in coincidence with the E$_x$ = 7230~keV state in any single measurement. Although the level scheme in $^{19}$F suggests a possible doublet structure of $J^\pi = \frac{3}{2}^+$ and $\frac{1}{2}^+$ in this region, further experimental data are required to clarify the level structure and confirm the presence of a $\frac{1}{2}^+$ state near this energy.

\textbf{33-35. E$_x$ = 7396$\pm$5 keV \& 7495$\pm$3 keV \& 7532$\pm$9 keV, $J^\pi = \dfrac{7}{2}^+ \& \dfrac{5}{2}^+ \& \dfrac{5}{2}^-$}

These states located at E$_x$ = 7396$\pm$5, 7495$\pm$3, and 7532$\pm$9~keV have spin-parity assignments of $J^\pi=\frac{7}{2}^+, \frac{5}{2}^+,$ and $\frac{5}{2}^-$, respectively. These levels lie well above the proton threshold and possess relatively high spins, suggesting that their direct contributions to the reaction rates under nova conditions are likely minor. Nonetheless, the resonance parameters have been experimentally measured and are summarized in Table~\ref{tab:Tab4}.

\textbf{36-37. E$_x$ = 7615$\pm$3 keV \& 7668$\pm$7 keV, $J^\pi = \dfrac{3}{2}^+$ \& $ \dfrac{3}{2}^-$}

The states located at E$_x$ = 7615$\pm$3 and 7668$\pm$7~keV are assigned as $J^\pi = \frac{3}{2}^+$ and $\frac{3}{2}^-$, respectively. The E$_x$ = 7668~keV state was previously interpreted as the mirror of the E$_x$ = 7702~keV state in $^{19}$F, which carries a spin-parity of $J^\pi=\frac{1}{2}^-$~\cite{2007Ne}. However, more recent experimental studies support a revised assignment of $J^\pi=\frac{3}{2}^-$, as summarized in Table~\ref{tab: pJ}.

\textbf{38-39. E$_x$ = 7764$\pm$3 keV \& 7872$\pm$19 keV, $J^\pi = \dfrac{3}{2}^+$ or $\dfrac{1}{2}^+$ \& $\dfrac{1}{2}^+$}

These two levels have been proposed as candidates for $l=0$ resonance states that may significantly influence the $^{18}$F($p, \alpha$)$^{15}$O reaction rate, particularly through interference with the subthreshold $J^\pi=\frac{1}{2}^+$ state. However, their spin-parity assignments remain uncertain, with inconsistent interpretations across the literature.

Kahl~\textit{et al.} \cite{2017Ka} suggests that the E$_x$ = 7790~keV state corresponds to $J^\pi=\frac{1}{2}^+$, whereas $J^\pi=\frac{1}{2}^+$ is assigned for the E$_x$ = 7872~keV level in Refs.~\cite{2012Mo,2009Da}. Especially, Mountford \textit{et al.} \cite{2012Mo} presents an R-matrix analysis that support assignment of $J^\pi=\frac{3}{2}^+$ and $\frac{1}{2}^+$ for the E$_x$ = 7764~keV and 7872~keV states, respectively. Consequently, the assignments of $J^\pi=\frac{3}{2}^+$ and $\frac{1}{2}^+$ are given weights of 0.75 and 0.25, respectively.

Considering the range of existing measurements, the present evaluation adopts the assignment of $J^\pi=\frac{1}{2}^+$ for the E$_x$ = 7872~keV state as the more plausible interpretation. Nevertheless, the E$_x$ = 7764~keV level may still play a role depending on its interference behavior. Given the possible role of these levels in interference effects involving other $\frac{1}{2}^+$ resonances, further experimental investigations are needed to clarify their properties and significance in reaction rate calculations.

\textbf{40. E$_x$ = 7995$\pm$5 keV, $J^\pi = \dfrac{5}{2}^+$ or $\dfrac{1}{2}^+$ or $\dfrac{5}{2}^-$}

The state has an excitation energy of E$_x$ = 7995$\pm$~5~keV. Its spin-parity assignment remains uncertain, with possible values reported as $J^\pi=\frac{5}{2}^+, \frac{1}{2}^+,$ and $\frac{5}{2}^-$, as summarized in Table~\ref{tab: pJ}. Among these, the assignment of $J^\pi = \frac{5}{2}^+$ is given the highest weight 0.5 in the present evaluation. The others are given weights of 0.25. This preference is primarily based on the analysis presented in Ref.~\cite{2012Mo}, which reported more detailed resolution of excitation levels above E$_x$ = 7600~keV and more comprehensive decay information than other studies~\cite{2009Da,2009Mu}. Additionally, the mirror analysis discussed in Ref.~\cite{2007Ne} favors this interpretation. Nevertheless, the possibility of alternative spin-parity values cannot be ruled out.
 

    \section{Conclusion}

We have presented a systematic and comprehensive Bayesian evaluation of the resonance parameters of excited states in $^{19}$Ne, focusing on astrophysically significant reactions: $^{15}$O($\alpha, \gamma$)$^{19}$Ne and $^{18}$F($p, \alpha$)$^{15}$O. The Bayesian methodology rigorously incorporates all available experimental data by reconstructing likelihood functions and combining them with physically motivated priors. This framework naturally accommodates asymmetric uncertainties and one-sided limits, which are challenging in conventional weighted-averaging techniques. The resulting posterior distributions yield well-defined resonance energies, branching ratios, and decay widths, together with reliable uncertainty estimates. Mirror symmetry analysis with the analog nucleus $^{19}$F was utilized to constrain parameters for states with insufficient direct experimental information, enabling a coherent and statistically robust evaluation.

The previous evaluation, Ref.~\cite{2007Ne}, focused exclusively on resonance levels relevant to the $^{18}$F($p, \alpha$)$^{15}$O reaction. Since then, around 20 further studies on this reaction have been published and are included in the present evaluation. Moreover, this work extends the scope of Ref.~\cite{2007Ne} by incorporating levels associated with the $^{15}$O($\alpha, \gamma$)$^{19}$Ne reaction. These new evaluations of $^{19}$Ne resonance parameters provide robust and critical input for calculations of thermonuclear reaction rates needed to model a trigger reaction for X-ray bursts and gamma-ray emission from classical novae. Thermonuclear reaction rates using these revised resonance parameters will be calculated, and their astrophysical impact for novae and X-ray bursts will be investigated in the following work. Future experimental efforts aimed at resolving remaining uncertainties-especially for unresolved multiplets and spin-parity ambiguities-will further strengthen the predictive power of astrophysical models.

\begin{acknowledgments}
This work was supported by the National Research Foundation of Korea (NRF) grants
funded by the Korea government (MSIT) (Grants No. RS-2024-00338255). This work was
also supported in part by the U.S. Dept. of Energy (DOE), Office of Science, Office of Nuclear
Physics under contract DE-AC05-00OR22725.
\end{acknowledgments}

\bibliographystyle{apsrev4-2}
\bibliography{ref}

\end{document}